\documentclass[twocolumn,resetfootnote]{aastex701}

\usepackage{natbib}
\usepackage{rotating}
\usepackage{xcolor}

\begin{document}

\title{\textit{Searching for GEMS:} Three warm Saturns and a super-Jupiter orbiting four early M-dwarfs  \footnote{Based on observations obtained with the Hobby-Eberly Telescope (HET), which is a joint project of the University of Texas at Austin, the Pennsylvania State University, Ludwig-Maximillians-Universitaet Muenchen, and Georg-August Universitaet Goettingen. The HET is named in honor of its principal benefactors, William P. Hobby and Robert E. Eberly}}

\author[0000-0001-5728-4735]{Pranav H. Premnath} \thanks{Corresponding author: Pranav Premnath}
\affiliation{Department of Physics \& Astronomy, University of California, Irvine, CA 92697, USA}
\email{premnatp@uci.edu}

\author[0000-0003-0149-9678]{Paul Robertson}
\affiliation{Department of Physics \& Astronomy, University of California, Irvine, CA 92697, USA}
\email{probert1@uci.edu}

\author[0000-0001-8401-4300]{Shubham Kanodia}
\affiliation{Earth and Planets Laboratory, Carnegie Institution for Science, 5241 Broad Branch Road, NW, Washington, DC 20015, USA}
\email{skanodia@carnegiescience.edu}

\author[0000-0003-4835-0619]{Caleb I. Ca\~nas}
\affiliation{Southeastern Universities Research Association, Washington, DC 20005, USA}
\affiliation{NASA Goddard Space Flight Center, 8800 Greenbelt Road, Greenbelt, MD 20771, USA}
\email{c.canas@nasa.gov}

\author[0000-0002-5463-9980]{Arvind F. Gupta}
\affiliation{U.S. National Science Foundation, National Optical-Infrared Astronomy Research Laboratory, 950 N. Cherry Ave., Tucson, AZ 85719, USA}
\email{arvind.gupta@noirlab.edu}

\author[0009-0009-4977-1010]{Michael Rodruck} 
\affiliation{Department of Physics, Engineering, and Astrophysics, Randolph-Macon College, Ashland, VA 23005, USA}
\email{mrodruck@gmail.com}

\author[0000-0002-7127-7643]{Te Han}
\affiliation{Department of Physics \& Astronomy, University of California, Irvine, CA 92697, USA}
\affiliation{Department of Physics and Kavli Institute for Astrophysics and Space Research, Massachusetts Institute of Technology, Cambridge, MA 02139, USA}
\email{tehan@mit.edu}

\author[0000-0002-4475-4176]{Henry A. Kobulnicky}
\affiliation{Department of Physics \& Astronomy, University of Wyoming, Laramie, WY 82070, USA}
\email{chipK@uwyo.edu}

\author[0000-0002-9082-6337]{Andrea S.J.\ Lin}
\affiliation{Department of Astronomy, California Institute of Technology, 1200 E California Blvd, Pasadena, CA 91125, USA}
\email{andrealin628@gmail.com}

\author[0000-0002-0048-2586]{Andrew Monson}
\affiliation{Steward Observatory, University of Arizona, 933 N.\ Cherry Ave, Tucson, AZ 85721, USA}
\email{monson.andy@gmail.com}

\author[]{Libby Allely} 
\affiliation{Pomona College, 333 N. College Way, Claremont, CA 91711, USA}
\email{lkal2024@mymail.pomona.edu}

\author[0009-0005-8787-8231]{Cooper Bailey} 
\affiliation{Department of Physics, Engineering, and Astrophysics, Randolph-Macon College, Ashland, VA 23005, USA}
\email{CooperBailey@go.rmc.edu}

\author[]{Alexina Birkholz} 
\affiliation{Department of Physics \& Astronomy, University of Wyoming, Laramie, WY 82070, USA}
\email{abirkho1@uwyo.edu}

\author[0009-0005-3725-5300]{Zack Beagle} 
\affiliation{Department of Physics, Engineering, and Astrophysics, Randolph-Macon College, Ashland, VA 23005, USA}
\email{ZackBeagle@go.rmc.edu}

\author[]{Philip Choi} 
\affiliation{Pomona College, 333 N. College Way, Claremont, CA 91711, USA}
\email{philip.Choi@pomona.edu}

\author[]{Nez Evans} 
\affiliation{Pomona College, 333 N. College Way, Claremont, CA 91711, USA}
\email{pre02012@mymail.pomona.edu}

\author[0000-0002-0885-7215]{Mark E. Everett} 
\affiliation{NSF NOIRLab, 950 N. Cherry Ave., Tucson, AZ 85719, USA}
\email{mark.everett@noirlab.edu}

\author[0009-0006-8374-089X]{Anna Fong} 
\affiliation{Maggie L. Walker Governor’s School, Richmond, VA 23220, USA}
\email{26afong@gsgis.k12.va.us}

\author[]{S. Nick Justice} 
\affiliation{Department of Physics \& Astronomy, University of Wyoming, Laramie, WY 82070, USA}
\email{sjustice@uwyo.edu}

\author[]{Ian Karfs} 
\affiliation{Department of Physics \& Astronomy, University of Wyoming, Laramie, WY 82070, USA}
\email{ikarfs@uwyo.edu}

\author[0000-0002-8623-8268]{Sage Santomenna} 
\affiliation{Pomona College, 333 N. College Way, Claremont, CA 91711, USA}
\email{mstp2022@mymail.Pomona.edu}

\author[]{Elsa Van Dyke} 
\affiliation{Pomona College, 333 N. College Way, Claremont, CA 91711, USA}
\email{ervh2024@mymail.pomona.edu}

\author[0009-0008-5739-9696]{Arissa Williams} 
\affiliation{Department of Physics, Engineering, and Astrophysics, Randolph-Macon College, Ashland, VA 23005, USA}
\email{ArissaWilliams@go.rmc.edu}

\author[0000-0003-4384-7220]{Chad F.\ Bender}
\affiliation{Steward Observatory, University of Arizona, 933 N.\ Cherry Ave, Tucson, AZ 85721, USA}
\email{cbender@arizona.edu}

\author[0000-0001-9662-3496]{William D. Cochran}
\affiliation{McDonald Observatory and Department of Astronomy, The University of Texas at Austin}
\affiliation{Center for Planetary Systems Habitability, The University of Texas at Austin}
\email{wdc@astro.as.utexas.edu}

\author[0000-0002-2144-0764]{Scott A.\ Diddams}
\affiliation{Electrical, Computer \& Energy Engineering, 440 UCB, University of Colorado, Boulder, CO 80309, USA}
\affiliation{Department of Physics, 390 UCB, University of Colorado, Boulder, CO 80309, USA}
\email{scott.diddams@colorado.edu}

\author[0000-0002-3853-7327]{Rachel B. Fernandes}
\altaffiliation{Center for Exoplanets and Habitable Worlds (CEHW) Fellow}
\affiliation{Department of Astronomy and Astrophysics, 525 Davey Laboratory, 251 Pollock Road, Penn State, University Park, PA, 16802, USA}
\affiliation{Center for Exoplanets and Habitable Worlds, 525 Davey Laboratory, 251 Pollock Road, Penn State, University Park, PA, 16802, USA}
\email{rbf5378@psu.edu}

\author[0000-0002-0078-5288]{Mark R.~Giovinazzi}
\affiliation{Department of Physics and Astronomy, Amherst College, 25 East Drive, Amherst, MA 01002, USA}
\email{mgiovinazzi@amherst.edu}

\author[0000-0003-1312-9391]{Samuel Halverson}
\affiliation{Jet Propulsion Laboratory, California Institute of Technology, 4800 Oak Grove Drive, Pasadena, California 91109}
\email{samuel.halverson@jpl.nasa.gov}

\author[0000-0001-9626-0613]{Daniel M.\ Krolikowski}
\affiliation{Steward Observatory, University of Arizona, 933 N.\ Cherry Ave, Tucson, AZ 85721, USA}
\email{krolikowski@arizona.edu}

\author[0000-0001-9596-7983]{Suvrath Mahadevan}
\affiliation{Department of Astronomy and Astrophysics, 525 Davey Laboratory, 251 Pollock Road, Penn State, University Park, PA, 16802, USA}
\affiliation{Center for Exoplanets and Habitable Worlds, 525 Davey Laboratory, 251 Pollock Road, Penn State, University Park, PA, 16802, USA}
\affiliation{Astrobiology Research Center, 525 Davey Laboratory, 251 Pollock Road, Penn State, University Park, PA, 16802, USA}
\email{suvrath@gmail.com}

\author[0000-0003-0241-8956]{Michael W.\ McElwain}
\affiliation{Exoplanets and Stellar Astrophysics Laboratory, NASA Goddard Space Flight Center, Greenbelt, MD 20771, USA} 
\email{michael.w.mcelwain@nasa.gov}

\author[0000-0001-8720-5612]{Joe P.\ Ninan}
\affiliation{Department of Astronomy and Astrophysics, Tata Institute of Fundamental Research, Homi Bhabha Road, Colaba, Mumbai 400005, India}
\email{indiajoe@gmail.com}

\author[0000-0001-8127-5775]{Arpita Roy}
\affiliation{Astrophysics \& Space Center, Schmidt Sciences, New York, NY 10011, USA}
\email{arpita308@gmail.com}

\author[0000-0001-7409-5688]{Gudmundur Stefansson}
\affiliation{Astrophysics \& Space Center, Schmidt Sciences, New York, NY 10011, USA}
\affiliation{Anton Pannekoek Institute for Astronomy, University of Amsterdam, Science Park 904, 1098 XH Amsterdam, The Netherlands}
\email{gstefansson@schmidtsciences.org}

\author[0000-0001-6160-5888]{Jason T. Wright}
\affiliation{Department of Astronomy and Astrophysics, 525 Davey Laboratory, 251 Pollock Road, Penn State, University Park, PA, 16802, USA}
\affiliation{Center for Exoplanets and Habitable Worlds, 525 Davey Laboratory, 251 Pollock Road, Penn State, University Park, PA, 16802, USA}
\affiliation{Penn State Extraterrestrial Intelligence Center, 525 Davey Laboratory, 251 Pollock Road, Penn State, University Park, PA, 16802, USA}
\email{astrowright@gmail.com}

\begin{abstract}
We report the confirmation and characterization of four transiting giant planets orbiting early-M dwarfs discovered by the \textit{Searching for Giant Exoplanets around M-dwarf Stars} (GEMS) survey: TOI-7189\,b, TOI-7265B\,b, TOI-7393\,b, and TOI-7394B\,b. Joint modeling of \textit{TESS} and ground-based photometry with precision radial velocities from the Habitable-zone Planet Finder and NEID spectrographs yields self-consistent orbital and physical parameters for all systems. The planets have short orbital periods ($P = 1.25-4.17$~days), masses spanning from $0.5\,M_{\rm J}$ to $2.1\,M_{\rm J}$, and radii comparable to Jupiter ($0.95\,R_{\rm J} < R_p < 1.02\,R_{\rm J}$). TOI-7189\,b ($0.50\,M_{\rm J}$), TOI-7265B\,b ($0.71\,M_{\rm J}$), and TOI-7393\,b ($0.61\,M_{\rm J}$) are Saturn-like in mass and density, whereas TOI-7394B\,b is a dense super-Jupiter ($2.10\,M_{\rm J}$, $\rho_p \approx 2.4$~g\,cm$^{-3}$) on a 1.25-day orbit. All hosts are early-M dwarfs with a narrow range of stellar properties, enabling a controlled comparison of giant-planet outcomes around low-mass stars. Three systems orbit super-solar metallicity stars, while TOI-7393 ($\mathrm{[Fe/H]} = -0.35 \pm 0.16$) is the most metal-poor GEMS host identified to date, and exhibits kinematics approaching the thin/thick-disk transition, suggestive of an older stellar population. Together, these systems reveal substantial diversity in the masses and bulk properties of short-period giant planets orbiting early-M dwarfs, demonstrating that markedly different planetary outcomes can arise around stars with otherwise similar fundamental properties.
\end{abstract}


\section{Introduction} \label{sec:intro}
Transiting planets around nearby M dwarf stars ($0.08M_{\odot}  \lesssim  M_{*} \lesssim 0.6M_{\odot}$; \citealt{pecaut2013}) offer a valuable window into planetary formation and evolution. M dwarfs are the most common stellar type in the Galaxy, comprising over 70\% of all stars \citep{henry2006}. Their smaller radii and masses make them ideal targets for detecting small planets through transits and radial velocities, and their long main-sequence lifetimes enable detailed studies of planetary evolution and atmospheric loss. 

The \textit{Transiting Exoplanet Survey Satellite} \citep[\textit{TESS};][] {ricker2014} has substantially expanded the known population of exoplanets orbiting M dwarfs through its nearly all-sky coverage and high-precision space-based photometry. In particular, \textit{TESS} has enabled the detection of short-period transiting planets around nearby low-mass stars, increasing the sample of well-characterized systems amenable to follow-up observations. Combined with precise astrometry from \textit{Gaia} \citep{gaia2023} and extensive ground-based spectroscopic and photometric observations, \textit{TESS} has enabled the discovery of planets across a broad range of stellar masses and orbital periods. Despite this progress, certain planet populations remain elusive, particularly \emph{giant exoplanets} ($R_p \gtrsim 8 R_\oplus$) around low-mass stars ($M_* \lesssim 0.6 M_\odot$). 

Empirically, the occurrence of gas giants decreases sharply with decreasing stellar mass \citep{johnson2010}. Surveys have shown that giant planets around M dwarfs are intrinsically rare across both short- and longer-period orbits. Short-period giant planets (periods $\lesssim$10 days) occur at rates below $\sim$0.1\% around M dwarfs \citep{gan2023, bryant2023, glusman2025}, while transiting Jupiter analogs are also uncommon in the low-mass stellar regime \citep{pass2023}. Most detections of giant planets around M-dwarfs to date have instead originated from radial velocity surveys of nearby M dwarfs \citep{endl2006, johnson2010, bonfils2013, maldonado2019, sabotta2021, schlecker2022, pinamoti2022, mignon2025}, which have similarly demonstrated that giant planets represent rare outcomes of planet formation around low-mass stars. This scarcity is consistent with expectations from core-accretion theory, where lower-mass protoplanetary disks associated with M dwarfs have reduced solid surface densities, leading to longer core-growth timescales at a given orbital radius and impeding the formation of massive cores before disk dispersal \citep{laughlin2004, alibert2005, burn2021}. However, the few systems that do host \emph{Giant Exoplanets around M-dwarf Stars (GEMS)} provide critical empirical anchors for testing models of planet formation and migration in the low-mass regime.

The \textit{Searching for GEMS} survey \citep[][]{kanodia2024a, glusman2025} has systematically identified and characterized gas giant planets orbiting low-mass stars using \textit{TESS} photometry combined with extensive ground-based observations. This work is also joined by other efforts to characterize GEMS, such as the GATOS \citep[Gas giAnts Transiting lOw-mass Stars;][]{Frensch2026} and MANGOS \cite[M dwarfs Accompanied by close-iN Giant Orbiters with SPECULOOS;][]{Triaud2023,Dransfield2026} surveys. Collectively, these efforts corroborate the idea that gas giant planets around M dwarfs are intrinsically rare, while also revealing a wide diversity in planetary radii, bulk densities, and orbital architectures. Several GEMS systems extend the population of known gas giant planets into the low-stellar-mass regime, enabling direct comparisons with analogous planets orbiting FGK stars and highlighting the value of stellar mass as an independent axis for testing planet formation models \citep[e.g.,][]{kanodia2025fgk}. As the GEMS sample continues to grow, it provides a critical foundation for comparative studies of giant planet demographics and physical properties across stellar types.

Population-level analyses within the GEMS program have begun to reveal emerging connections between giant planets and the properties of their low-mass host stars. Studies of the growing GEMS sample suggest that giant planets around M dwarfs are often found around stars with higher inferred metallicities \citep{gan2023, han2024}, although these trends remain tentative given the uncertainties associated with M-dwarf metallicity determinations. This behavior is qualitatively consistent with well-established giant planet--metallicity correlation first identified for FGK dwarfs \citep[e.g.,][]{gonzalez1997, santos2004, fischer2005}, in which stars with higher metallicities are significantly more likely to host gas giant planets. The correlation is broadly consistent with expectations from the core-accretion paradigm, wherein higher-metallicity protoplanetary disks contain larger solid reservoirs that facilitate the formation of massive planetary cores capable of accreting substantial gaseous envelopes \citep{alibert2017}. These results underscore the importance of precise stellar characterization for interpreting individual systems within a broader demographic context.

In parallel with the characterization of individual systems, the GEMS program has begun to place these discoveries in a broader population context. Current surveys indicate that giant planets around M dwarfs are intrinsically rare, but the relatively small number of confirmed systems limits the statistical precision with which their demographic properties can be quantified \citep{gan2023, bryant2023, glusman2025}. While emerging samples have begun to suggest possible trends---such as a dependence on host-star metallicity or stellar mass---these remain tentative given the limited statistics and uncertainties in stellar parameters. Expanding the sample of well-characterized systems will enable more robust tests of these potential correlations, including whether giant planet occurrence varies across the M-dwarf sequence, how it depends on host-star metallicity, and whether the mass distribution differs from that observed around FGK stars. Such efforts are essential for providing empirical constraints on models of planet formation and migration in the M-dwarf regime.

Motivated by these limitations, the GEMS program was designed from the outset to conduct a 200~pc \citep{kanodia2025}, volume-limited survey of low-mass stars, with the initial 100~pc sample \citep{glusman2025} serving as a pilot study to validate the survey methodology and establish baseline occurrence constraints. The full 200~pc survey increases the number of monitored low-mass stars by roughly an order of magnitude, substantially improving sensitivity to these intrinsically rare planetary systems. This expanded sample is designed to better probe the extreme tail of the planet population---systems that are expected to form only under the most favorable disk and host-star conditions---and to place stronger constraints on the occurrence and properties of giant planets around low-mass stars.

Here, we present the confirmation and characterization of four short-period transiting giant planets---TOI-7189\,b, TOI-7265B\,b, TOI-7393\,b, and TOI-7394B\,b. All four systems host gas-rich planets with planetary radii  $\sim$10 $R_\oplus$ and orbital periods shorter than 5~days, placing them among the largest planets in compact orbits known to transit M-dwarfs. The planets span a wide mass range of approximately $0.4$--$2.1 M_{\rm J}$, while orbiting a narrow range of early-M host stars (M0--M2) representative of the majority of GEMS hosts. Their short orbital periods and nearby hosts make these systems particularly well suited for precise radial velocity follow-up and future atmospheric characterization. Collectively, these four systems provide a valuable comparative sample for probing the formation, internal structure, and evolutionary pathways of giant planets around low-mass stars, and for assessing the role of host-star properties---particularly metallicity---in shaping their final architectures.

This paper is structured as follows. Section~\ref{sec:obs} describes the observations and data reduction, including the \textit{TESS} light curves, ground-based photometry, radial velocities, and high-resolution imaging. Section~\ref{sec:stellar} presents the stellar characterization of all four host stars. Section~\ref{sec:modeling} outlines the joint modeling of the transit and radial velocity data. In Section~\ref{sec:discussion}, we place these systems in the broader context of the GEMS sample and discuss implications for planet formation and evolution around low-mass stars. Finally, Section~\ref{sec:conclusion} summarizes our findings and discusses prospects for future planet formation studies.

\section{Observations} \label{sec:obs}

\subsection{TESS Photometry} \label{sec:tess}

All four of TOI-7189\,b, TOI-7265B\,b, TOI-7393\,b, and TOI-7394B\,b were identified as transiting planet candidates in \textit{TESS} Quick-Look Pipeline (QLP) light curves as part of the Faint Star Search \citep{kunimoto2022a}. TOI-7393\,b and TOI-7394B\,b were additionally identified in earlier work by \citet{montalto2023}. These targets were also independently recovered and vetted within our 200 pc GEMS survey, which employs \texttt{TESS-Gaia Light Curve} \citep[\texttt{TGLC};][]{han2023} photometry and a uniform, multi-stage pipeline combining automated transit searches, visual inspection, and Bayesian light curve modeling. The targets were observed across multiple \textit{TESS} sectors during both the primary and extended missions, yielding multi-epoch photometry well suited for transit detection, ephemeris refinement, and long-term consistency checks (Table~\ref{tab:tess_obs}; \citet{tessmast}).

\begin{table*}
\centering
\caption{\textit{TESS} photometric observations of the four systems.}
\label{tab:tess_obs}
\begin{tabular}{lll}
\hline\hline
Target & Observed Sectors & Time Baseline \\
\hline

TOI-7189 & 80 & 2024 Jun 18 -- 2024 Jul 15 \\

TOI-7265B &
15, 16, 55, 56, 75, 76, 82, 83 &
2019 Aug -- 2024 Oct \\

TOI-7393 &
16, 22, 23, 24, 49, 50, 76, 77 &
2019 Sep -- 2024 Apr \\

TOI-7394B &
15, 16, 22, 24, 49, 50, 51, 76, 77, 78, 82 &
2019 Aug -- 2024 Sep \\

\hline
\end{tabular}
\end{table*}

We adopted light curves extracted with \texttt{TGLC} \citep{tglcmast}, which performs point-spread function modeling informed by \textit{Gaia} DR3 astrometry to deblend stellar flux contributions within the \textit{TESS} pixels, making it particularly effective for faint targets and moderately crowded fields. This approach yields improved photometric precision and accurate transit depths by explicitly accounting for flux contamination from nearby sources \citep{han2025}.

We detrended the raw \texttt{TGLC} light curves ($\texttt{aperture\_flux}$) using the \texttt{wotan} package \citep{hipke2019}, as implemented in the \texttt{TESS-miner} pipeline described by \citet{glusman2025}. The detrending employs a cosine filtering approach to remove long-timescale stellar and instrumental variability while preserving signals on transit timescales. Transit windows were masked during the detrending procedure to prevent distortion of the transit profiles. The resulting light curves show reduced low-frequency variability compared to the input calibrated photometry and are suitable for joint transit modeling across multiple sectors. 

The differing photometric uncertainties between sectors are primarily driven by the varying \textit{TESS} observing cadences adopted throughout the mission and propagated through the \texttt{TGLC} light curve extraction. In the nominal \textit{TESS} mission (Sectors 1--26), full-frame images were obtained at 1800 s cadence, which was reduced to 600 s during the first extended mission (Sectors 27--55) and to 200 s beginning in Sector 56. The shorter integration times generally produce larger point-to-point scatter because each exposure contains fewer collected photons, leading to systematically larger median absolute deviations in the extracted light curves. As a result, sectors observed at shorter cadence typically exhibit larger apparent photometric uncertainties.

\begin{figure*}
    \centering
    \includegraphics[width=\textwidth]{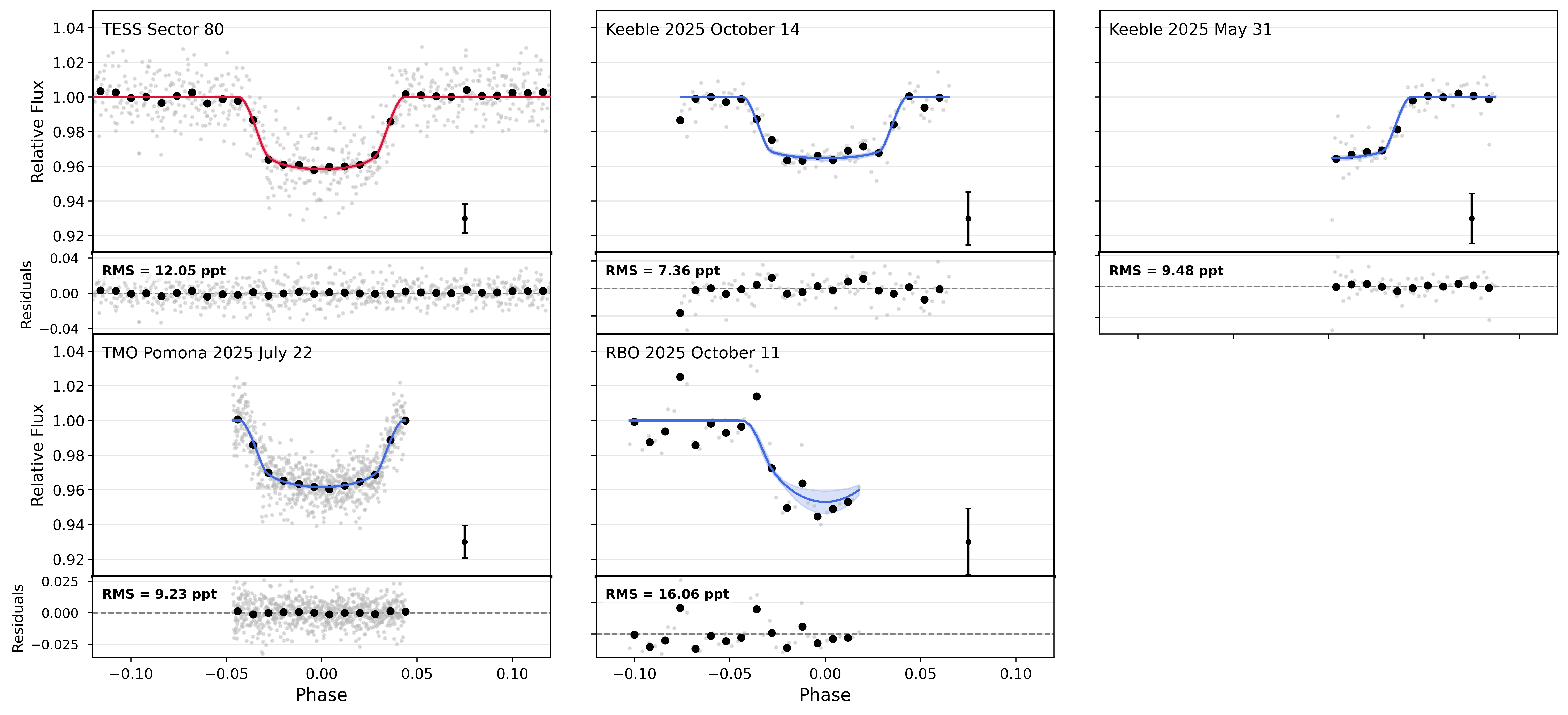}
    \caption{Phase-folded light curves of the host star TOI-7189, folded to the orbital period of TOI-7189\,b, from \textit{TESS} (red) and ground-based observations (blue). Gray points show the detrended photometry and solid black circles show inverse-variance weighted phase-binned data for visual clarity. Solid curves show the best-fit transit models from the joint analysis, with shaded regions indicating the 68\% credible intervals. Ground-based data were modeled with fixed dilution, while \textit{TESS} data were fit with floating dilution. Residuals and their RMS scatter are shown in the lower panels, with binned residuals overplotted in black. Representative median absolute deviations of the photometric measurements are indicated by the error bar in each panel.}
    \label{fig:7189_lc}
\end{figure*}

\begin{figure*}
    \centering
    \includegraphics[width=\textwidth]{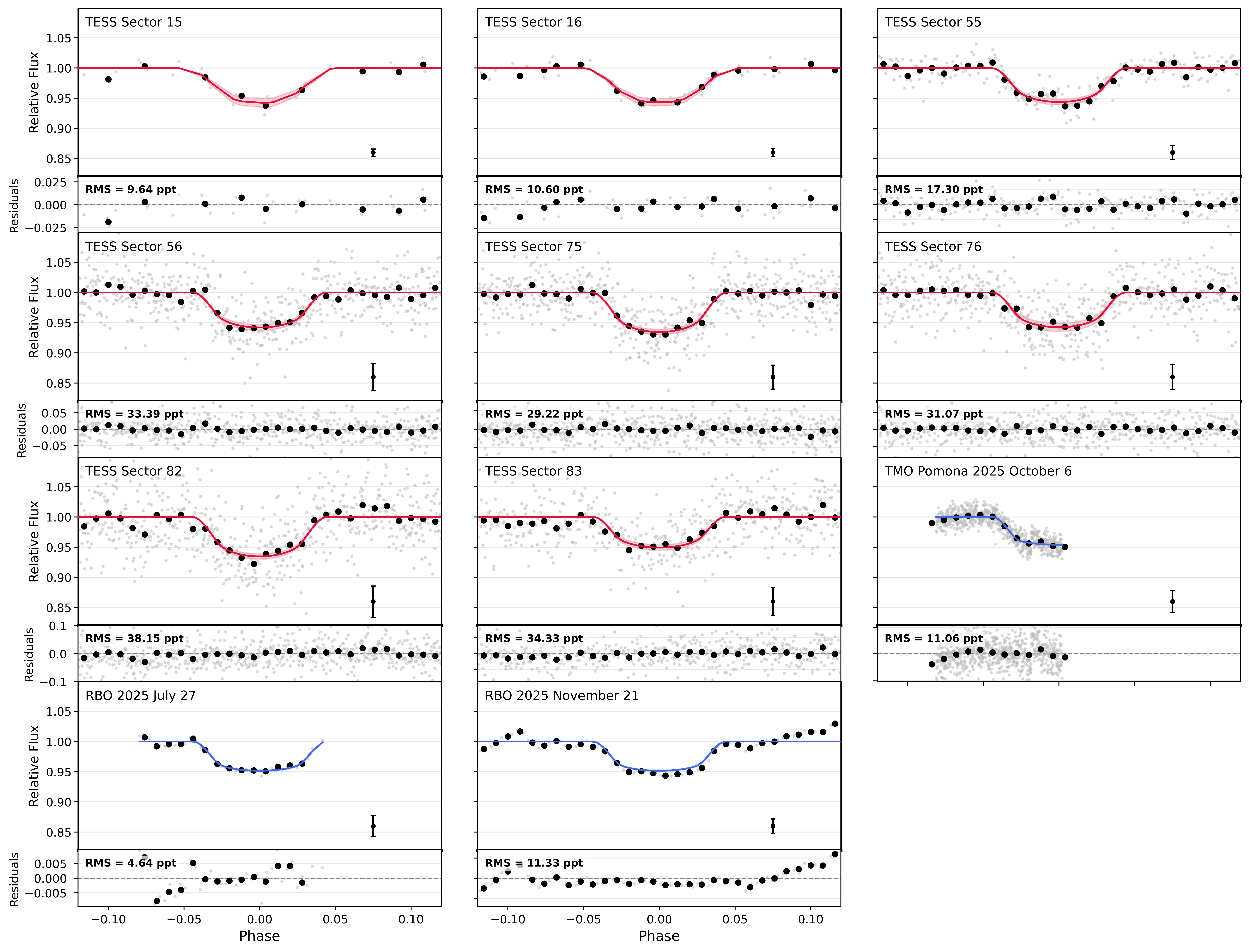}
    \caption{
    Same as Figure~\ref{fig:7189_lc}, for \textbf{TOI-7265B}}
    \label{fig:7265_lc}
\end{figure*}

\begin{figure*}
    \centering
    \includegraphics[width=\textwidth]{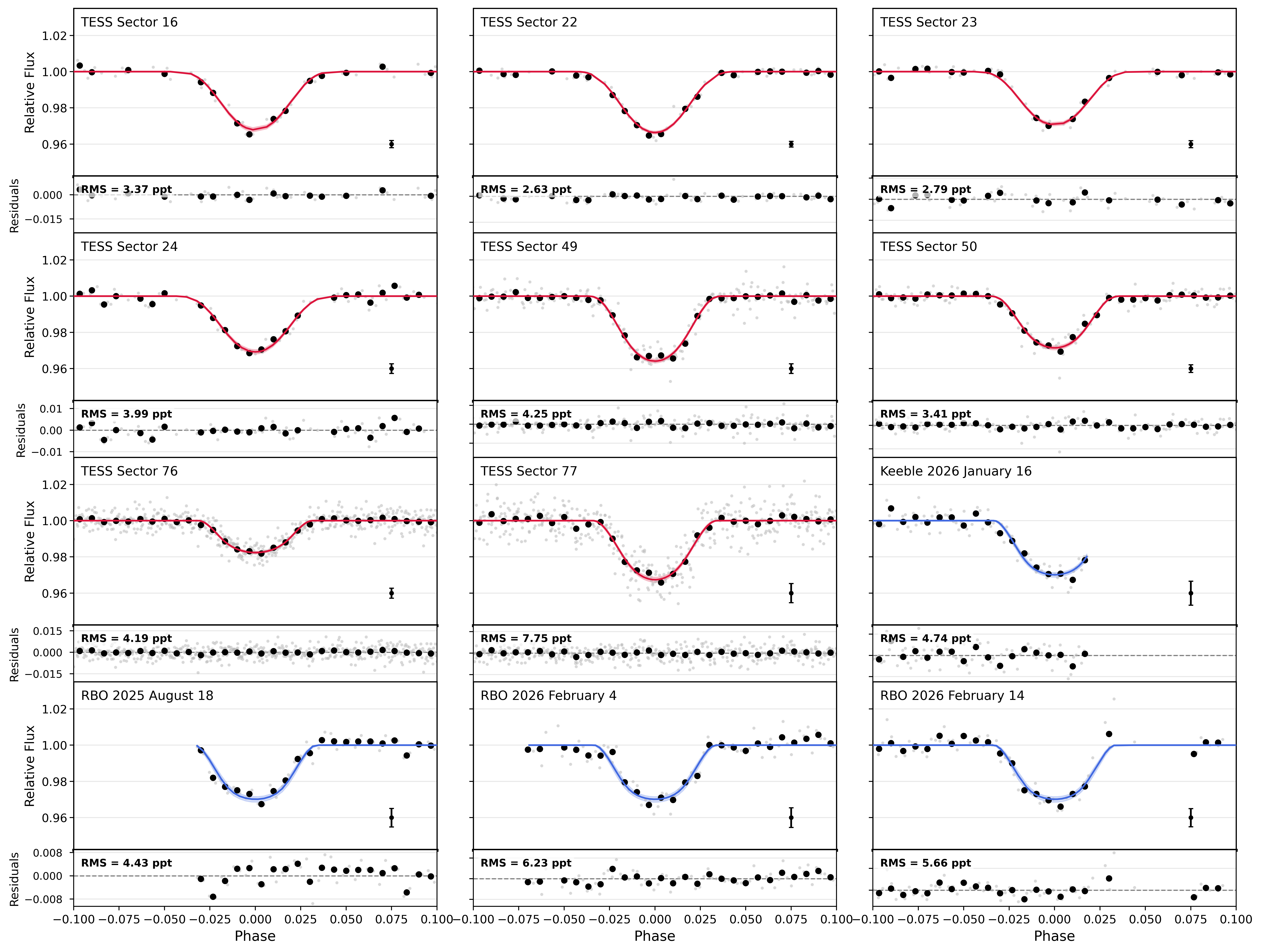}
    \caption{
    Same as Figure~\ref{fig:7189_lc}, for \textbf{TOI-7393}}
    \label{fig:7393_lc}
\end{figure*}

\begin{figure*}
    \centering
    \includegraphics[width=\textwidth]{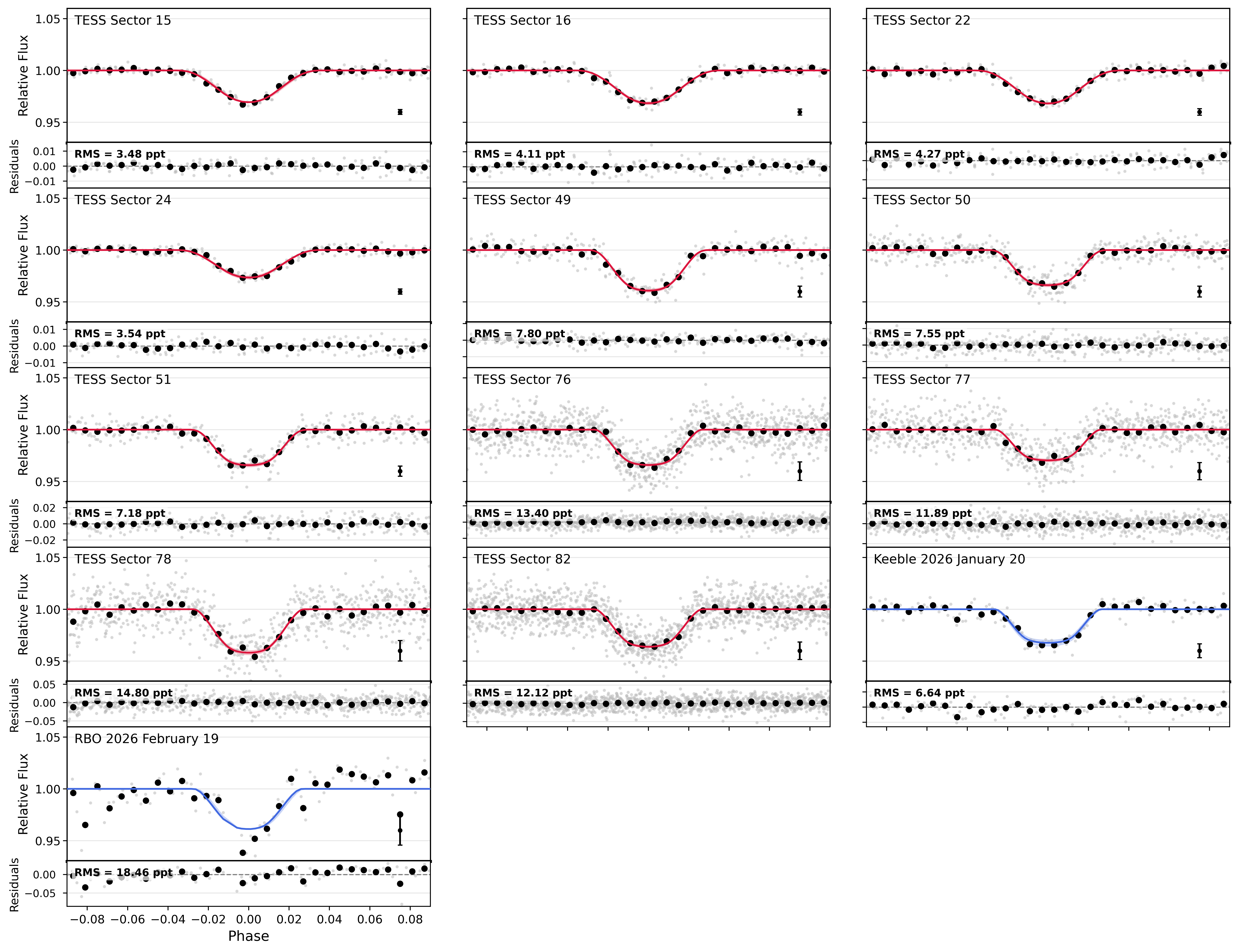}
    \caption{
    Same as Figure~\ref{fig:7189_lc}, for \textbf{TOI-7394B}}
    \label{fig:7394_lc}
\end{figure*}

Visual inspection of each sector confirms coherent, periodic transit signals with consistent depths and durations across all available data. The final detrended and phase-folded \textit{TESS} light curves are shown in Figure~\ref{fig:7189_lc} for TOI-7189\,b, Figure~\ref{fig:7265_lc} for TOI-7265B\,b, Figure~\ref{fig:7393_lc} for TOI-7393\,b, and Figure~\ref{fig:7394_lc} for TOI-7394B\,b. The panels showing the best-fit transit models (red curves for \textit{TESS}) form the basis of our transit analysis.

\subsection{Ground-based Photometric Observations} \label{sec:ground}

\textit{TESS} full-frame images have a coarse pixel scale ($21''$\,pix$^{-1}$) and are therefore susceptible to blending and flux contamination from nearby sources. Although the \texttt{TGLC} pipeline provides a PSF-based correction for crowding using Gaia catalog information, residual dilution from unresolved or poorly characterized sources may remain. To independently validate the transit signals, confirm their association with the target stars, and provide high-angular-resolution transit depth measurements free from blending assumptions, we obtained extensive ground-based time-series photometric observations of TOI-7189, TOI-7265B, TOI-7393, and TOI-7394B. A summary of all ground-based photometric follow-up observations is listed in Table~\ref{tab:ground_followup}.

\begin{table*}[ht!]
\centering
\caption{Ground-based photometric follow-up observations.}
\label{tab:ground_followup}
\begin{tabular}{lllccll}
\hline\hline
Target & Date (UT) & Observatory & Duration Observed (hr) & Exposure Time (s) & Filter & Defocused \\
\hline

TOI-7189 & 2025 May 31 & Keeble Observatory (0.4\,m) & 2.2 & 120 & $i$ & No \\

TOI-7189 & 2025 Jul 22 & Table Mountain Observatory (1\,m) & 2.0 & 10 & $i$ & Yes \\

TOI-7189 & 2025 Oct 11 & Red Buttes Observatory (0.6\,m) & 3.0 & 240 & Bessell $I$ & No \\

TOI-7189 & 2025 Oct 14 & Keeble Observatory (0.4\,m) & 3.4 & 120 & $i$ & No \\

TOI-7265B & 2025 Jul 27 & Red Buttes Observatory (0.6\,m) & 6.25 & 240 & Bessell $I$ & No \\

TOI-7265B & 2025 Oct 06 & Table Mountain Observatory (1\,m) & 2.0 & 10 & $i$ & Yes \\

TOI-7265B & 2025 Nov 21 & Red Buttes Observatory (0.6\,m) & 6.1 & 240 & Bessell $I$ & No \\

TOI-7393 & 2025 Aug 18 & Red Buttes Observatory (0.6\,m) & 4.8 & 240 & Bessell $I$ & No \\

TOI-7393 & 2026 Jan 16 & Keeble Observatory (0.4\,m) & 3.2 & 120 & $i$ & No \\

TOI-7393 & 2026 Feb 04 & Red Buttes Observatory (0.6\,m) & 6.0 & 150 & Bessell $I$ & No \\

TOI-7393 & 2026 Feb 14 & Red Buttes Observatory (0.6\,m) & 6.7 & 150 & Bessell $I$ & No \\

TOI-7394B & 2026 Jan 20 & Keeble Observatory (0.4\,m) & 4.8 & 120 & $i$ & No \\

TOI-7394B & 2026 Feb 19 & Red Buttes Observatory (0.6\,m) & 6.0 & 150 & Bessell $I$ & No \\

\hline\hline
\end{tabular}
\end{table*}

\subsubsection{Red Buttes Observatory} \label{sec:rbo}
We obtained time-series photometric observations of TOI-7189, TOI-7265B, TOI-7393, and TOI-7394B using the 0.6\,m University of Wyoming Red Buttes Observatory (RBO) telescope near Laramie, Wyoming, USA \citep{kasper2016}. TOI-7265B was observed on UT~2025 July 27 and UT~2025 November 21 using 240\,s exposures in the Bessell $I$ band. TOI-7189 was observed on UT~2025 October 11 with 240\,s exposures in the same band. TOI-7393 was observed on UT~2025 August 18 using 240\,s exposures, and on UT~2026 February 04 and UT~2026 February 14 using 150\,s exposures, in the Bessell $I$ band. TOI-7394B was observed on UT~2026 February 19 using 150\,s exposures, in the Bessell $I$ band.

The observations were obtained using a thermoelectrically cooled e2V CCD47-10 1024$\times$1024 detector (operated at $-30^\circ$C in July and $-40^\circ$C otherwise) with a pixel size of 13~$\mu$m, yielding a plate scale of 0.52\arcsec\,pixel$^{-1}$ and a field of view of 9.0\arcmin.

The raw images were reduced using a custom Python differential aperture photometry pipeline described in \citet{kanodia2024b}. The processing included bias and dark subtraction followed by division by a median-combined, normalized dome flat. Instrumental systematics were corrected by masking the transit events and removing low-order trends in the out-of-transit baseline when necessary. 

For TOI-7189, TOI-7265B, and TOI-7393, differential aperture photometry was extracted by iterating over ensembles of comparison stars and aperture sizes to minimize the out-of-transit root-mean-square (RMS) scatter of the light curves. TOI-7265B has a wide ($\sim$18\arcsec) companion (TOI-7265A); we verified that the transit signal is associated with the fainter target star (TOI-7265B) and not the brighter companion. This is discussed further in Section~\ref{sec:imaging}, and Table Mountain Observatory photometry confirms TOI-7265B as the transit host star (Section~\ref{sec:pomona}).

TOI-7394B has a nearby stellar neighbor (TOI-7137, or TOI-7394A) within the photometric field ($\sim$4\arcsec), which introduced contamination in aperture photometry. For this target, point-spread-function (PSF) photometry was therefore adopted to better separate the flux contributions of the two sources. The resulting light curve shows a transit signal on TOI-7394B phased consistently with the \textit{TESS} ephemeris. Ground-based photometry obtained with the Keeble Observatory (Section~\ref{sec:keeble}) further confirmed TOI-7394B as the transit host.

\subsubsection{Pomona College 1\,m Telescope at Table Mountain Observatory} \label{sec:pomona}

Additional ground-based photometric observations were obtained with the 1\,m Pomona College telescope at Table Mountain Observatory, San Bernardino County, California, USA. Observations were obtained using a Photometrics Prime 95B back-illuminated sCMOS camera, with a pixel scale of 0.225\arcsec\,pix$^{-1}$ and a field of view of 6\arcmin $\times$ 6\arcmin. 
TOI-7189 was observed on UT~2025 July 22 and TOI-7265B on UT~2025 October 06 using 10\,s exposures in the Sloan $i$ band. Each observing sequence spanned approximately 2 hours under clear conditions and fully covered the predicted transit events. The telescope was defocused to broaden the stellar PSF and improve photometric stability. 

The Pomona data were reduced using standard image calibration procedures, including bias subtraction and flat-fielding. Time-series differential photometry was extracted using \texttt{AstroImageJ} \citep{collins2016}.  Circular apertures of radii 5.13\arcsec and 2.71\arcsec respectively (1.4 x FWHM) were used, with the sky background estimated from annuli extending from 11.25\arcsec to 18\arcsec (50 to 80 pixels). The resulting light curves were normalized and timestamps were converted to BJD$_\mathrm{TDB}$, using the \texttt{barycorrpy} \citep{kanodia2018} package, which implements the algorithms described by \citet{wright2014}.

\subsubsection{Keeble Observatory} \label{sec:keeble}
TOI-7189, TOI-7393, and TOI-7394B were additionally observed with the 0.4 m Keeble Observatory telescope operated by Randolph-Macon College, Ashland, Virginia, USA. The telescope is a $f$/8 Ritchey-Chr\'{e}tien manufactured by ASA. 
The observations at Keeble were done using a SBIG Aluma CCD47-10 camera with a gain of 1.22 e- /ADU, and a plate scale of 0.83\arcsec\,pix$^{-1}$, resulting in a field of view of 14.1\arcmin $\times$ 14.1\arcmin. All observations were completed using the Sloan $i$-band, with 120 s exposures. 

Time-series photometry for TOI-7189 was obtained on UT 2025 May 31 and UT 2025 October 14. The target was observed on UT 2025 October 14 with average seeing conditions of 3.8\arcsec and on UT 2025 May 31 with average seeing conditions of 5.3\arcsec. TOI-7393 was observed on UT 2026 January 16, with an average seeing of 4.3\arcsec. TOI-7394B was observed on UT 2026 January 20, with an average seeing of 2.5\arcsec. Differential aperture photometry was performed using \texttt{AstroImageJ} \citep{collins2016} for TOI-7189 and TOI-7393. Photometry for TOI-7394B was performed using PSF photometry, due to the presence of TOI-7394A, using the differential photometry pipeline described in \citet{kanodia2024b}.

The resulting ground-based light curves from all facilities are consistent with the \textit{TESS} ephemerides for all four systems and show no evidence for offset transit events or nearby eclipsing binaries. Because the ground-based observations use substantially smaller photometric apertures than \textit{TESS}, the detected transit signals can be spatially localized to the target stars, confirming them as the true hosts of the transiting planets. The measured transit depths are consistent with the dilution-corrected \texttt{TGLC} values within the uncertainties, demonstrating the reliability of the \texttt{TGLC} contamination correction. These observations therefore confirm that the transit signals originate from the target stars and provide strong constraints on photometric dilution in the \textit{TESS} data.

\subsection{Radial Velocities} \label{sec:rvs}

We obtained radial velocity (RV) observations of TOI-7189, TOI-7265B, TOI-7393, and TOI-7394B to confirm the planetary nature of the transiting signals and to measure the companion masses. RV data were acquired using the Habitable-zone Planet Finder (HPF; \citealt{mahadevan2012, mahadevan2014}), a stabilized \citep{stefansson2016}, fiber-fed \citep{kanodia2018b}, near-infrared spectrograph on the 10\,m Hobby--Eberly Telescope (HET) at McDonald Observatory \citep{ramsey1998, hill2021}. We also acquired data using NEID, a fiber-fed environmentally stabilized spectrograph mounted on the WIYN 3.5\,m telescope\footnote{The WIYN Observatory is a joint facility of the NSF’s National Optical-Infrared Astronomy Research Laboratory, Indiana University, the University of Wisconsin-Madison, Pennsylvania State University, Purdue University and Princeton University.}\citep{halverson2016, schwab2016, robertson2019}. HPF operates at a resolving power of $R\sim55{,}000$ over the wavelength range 0.81--1.28~$\mu$m and is optimized for precision RV measurements of low-mass M-dwarf stars, while NEID provides high-resolution ($R\sim110{,}000$) red-optical spectra for precise Doppler measurements at a wavelength range 380--930 nm.

\begin{figure*}
    \centering
    \includegraphics[width=0.48\textwidth]{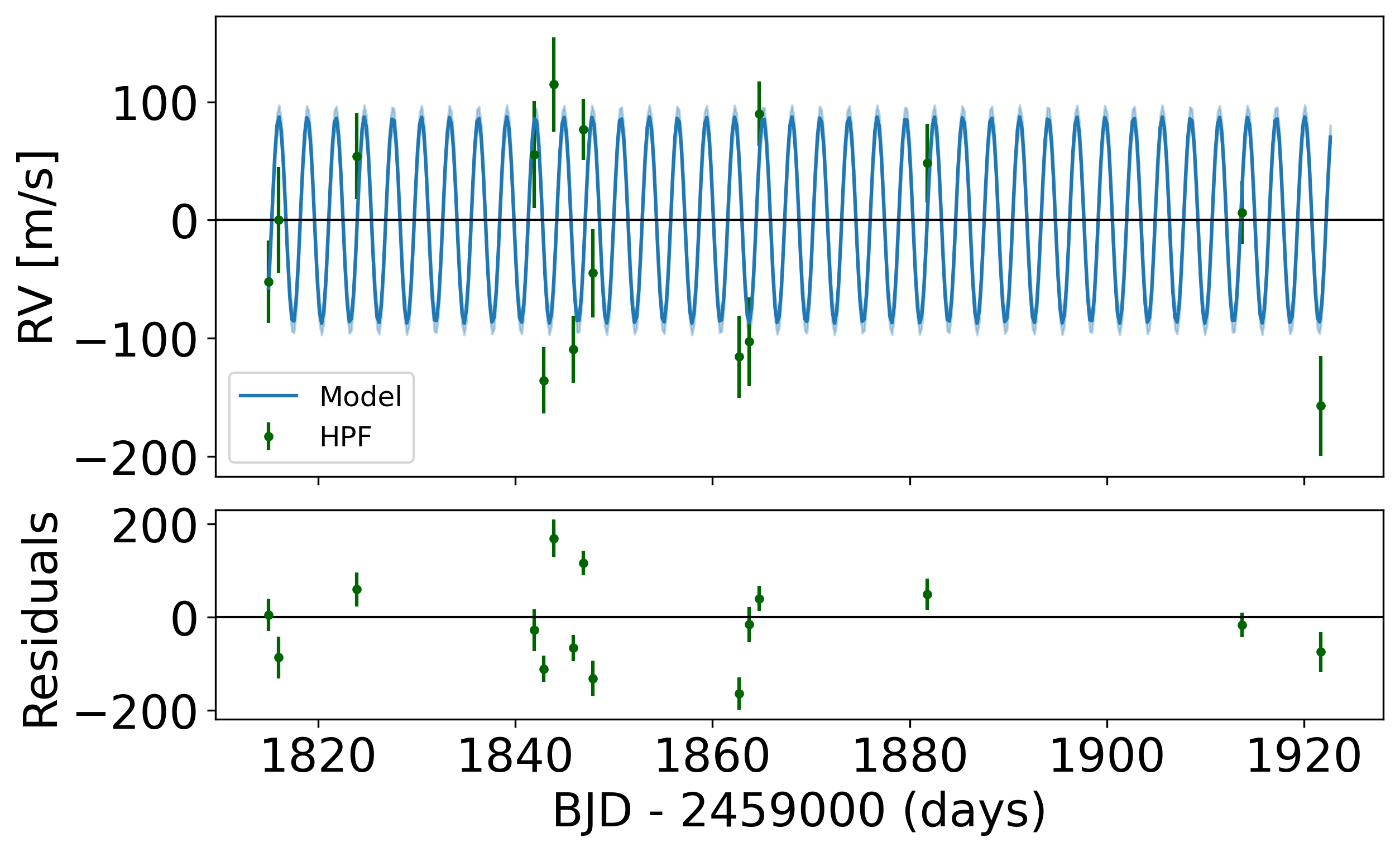}
    \hfill
    \includegraphics[width=0.48\textwidth]{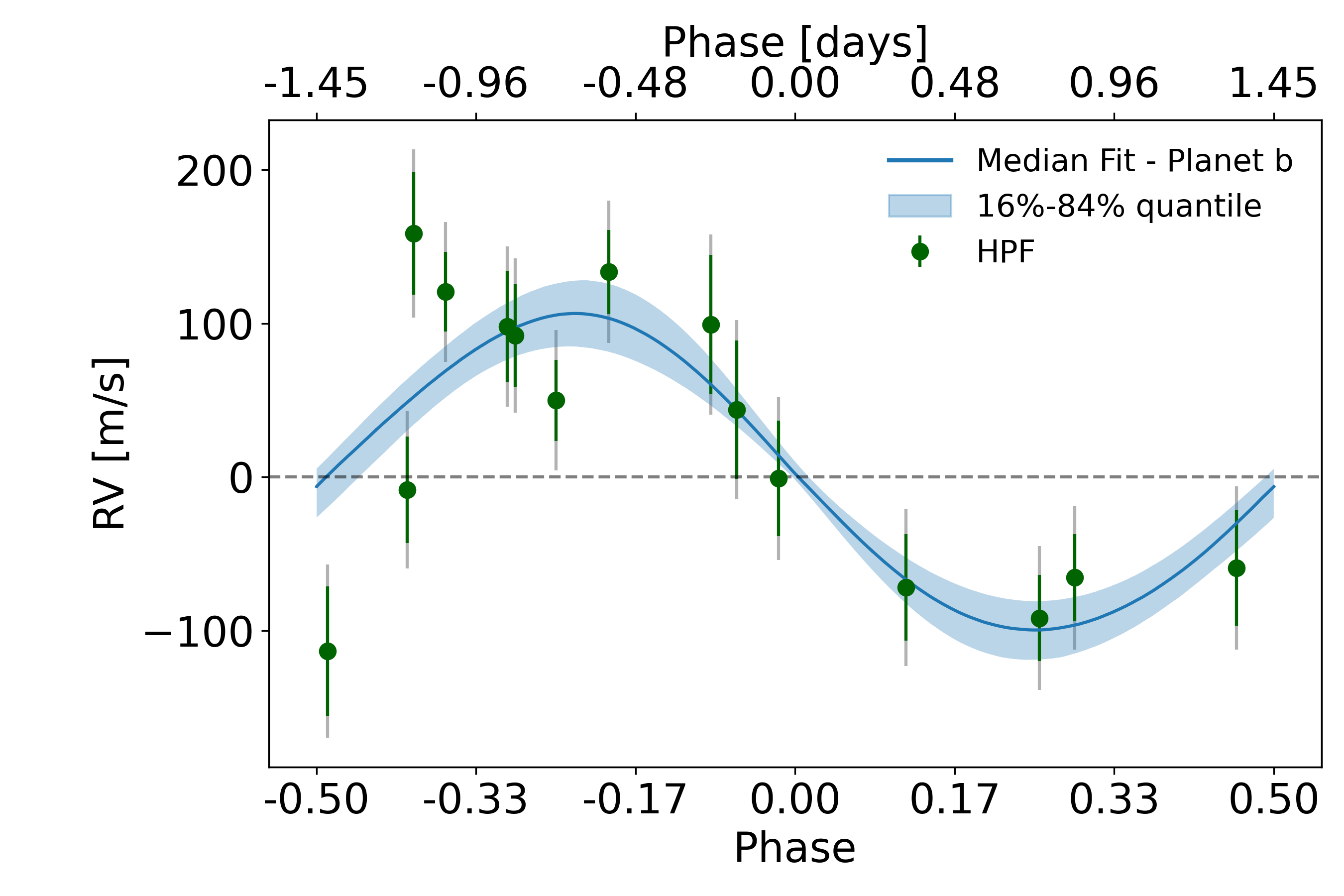}
    \caption{
    Radial velocity observations of \textbf{TOI-7189}.
    \emph{Left:} RV time-series obtained with HPF, with the best-fit Keplerian model overplotted.
    \emph{Right:} RVs phase-folded to the orbital period of TOI-7189\,b. The green error bars indicate the internal instrumental uncertainties, while the gray error bars represent the total uncertainties after adding the fitted RV jitter term in quadrature. We note that the jitter is minimal and not easily visible.
    The solid curve shows the median posterior model, and the shaded region denotes the 16-84\% quantile. The observed semi-amplitude confirms the planetary mass inferred from the joint photometric and RV analysis.
    }
    \label{fig:rv_7189}
\end{figure*}

\begin{figure*}
    \centering
    \includegraphics[width=0.48\textwidth]{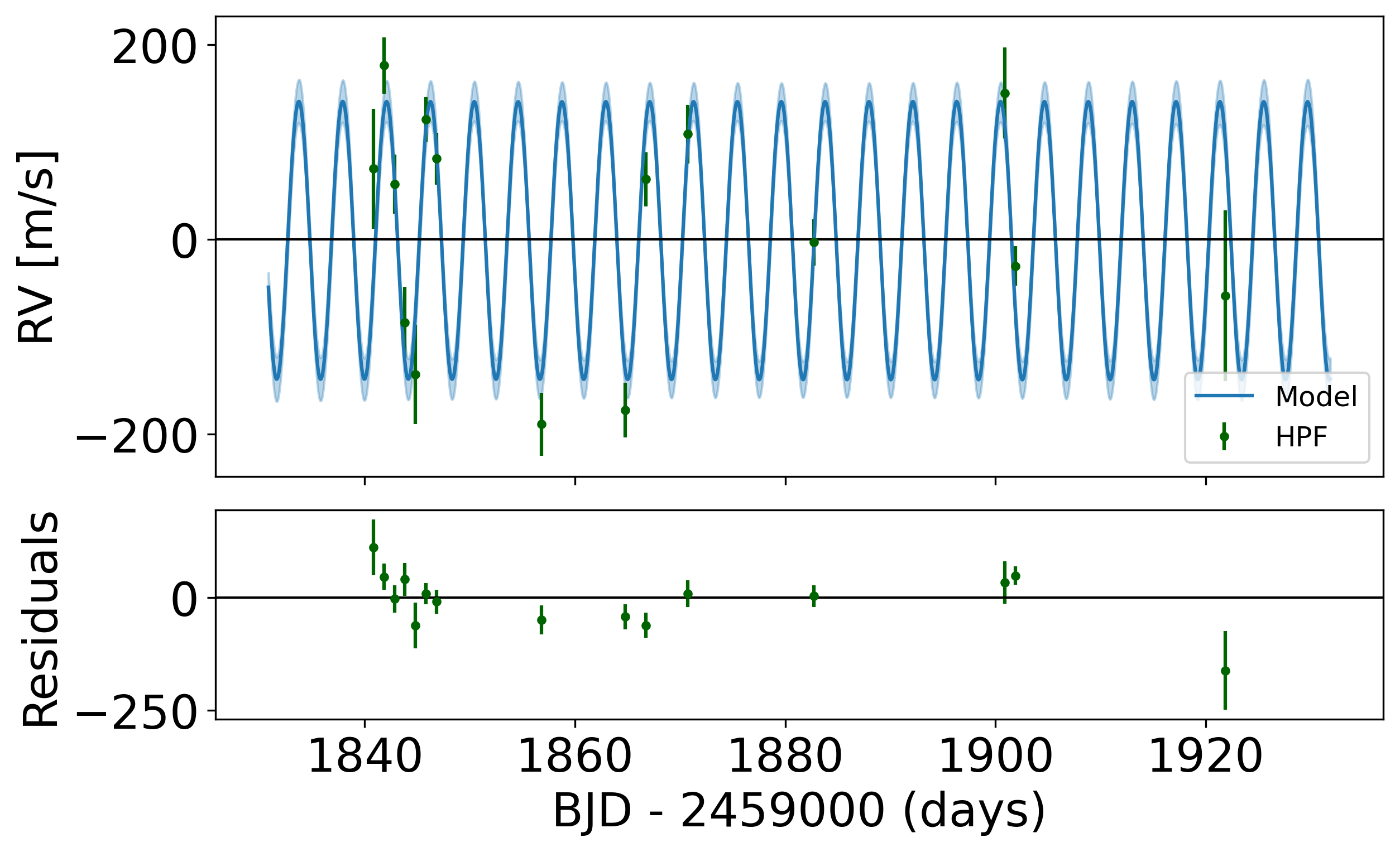}
    \hfill
    \includegraphics[width=0.48\textwidth]{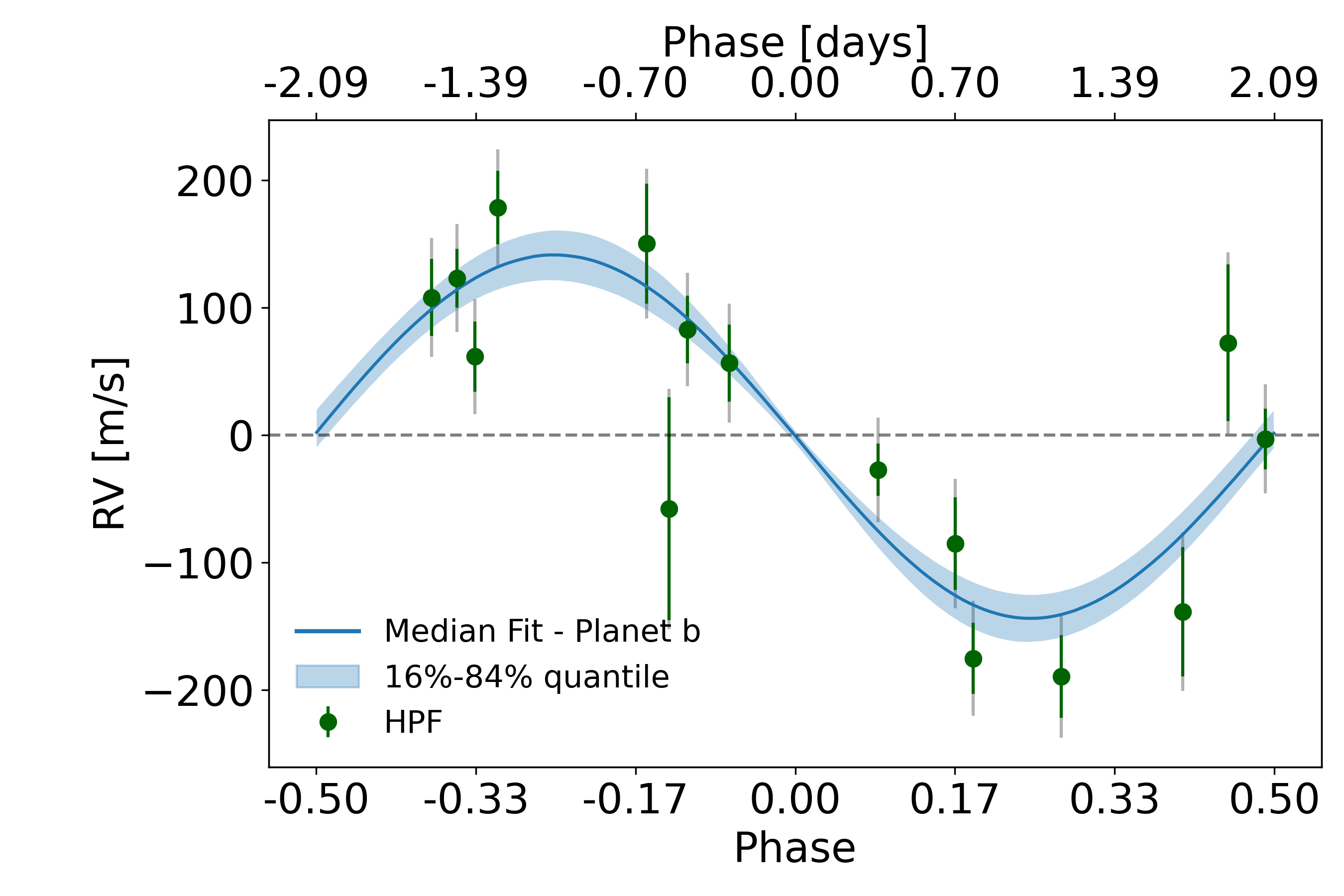}
    \caption{
    Radial velocity observations of \textbf{TOI-7265B}.
    \emph{Left:} HPF RV time-series with the best-fit Keplerian model.
    \emph{Right:} Phase-folded RVs for TOI-7265B\,b.
    }
    \label{fig:rv_7265}
\end{figure*}

For TOI--7189 and TOI--7265B, we obtained a total of 17 HPF observing visits for each target spanning approximately 80 days. During each visit, two consecutive exposures of 945~s were acquired (34 individual exposures in total per target). The paired exposures from each night were subsequently combined to produce a single RV measurement per visit. Measurements with low signal-to-noise ratios (SNR), poor wavelength-solution fits, or clear outliers relative to the time-series were excluded, leaving 15 high-quality RV measurements for each system. The resulting radial velocities exhibit coherent variations in phase with the photometric ephemerides (Figures~\ref{fig:rv_7189} and~\ref{fig:rv_7265}).

\begin{figure*}[t]
    \centering
    \includegraphics[width=0.48\textwidth]{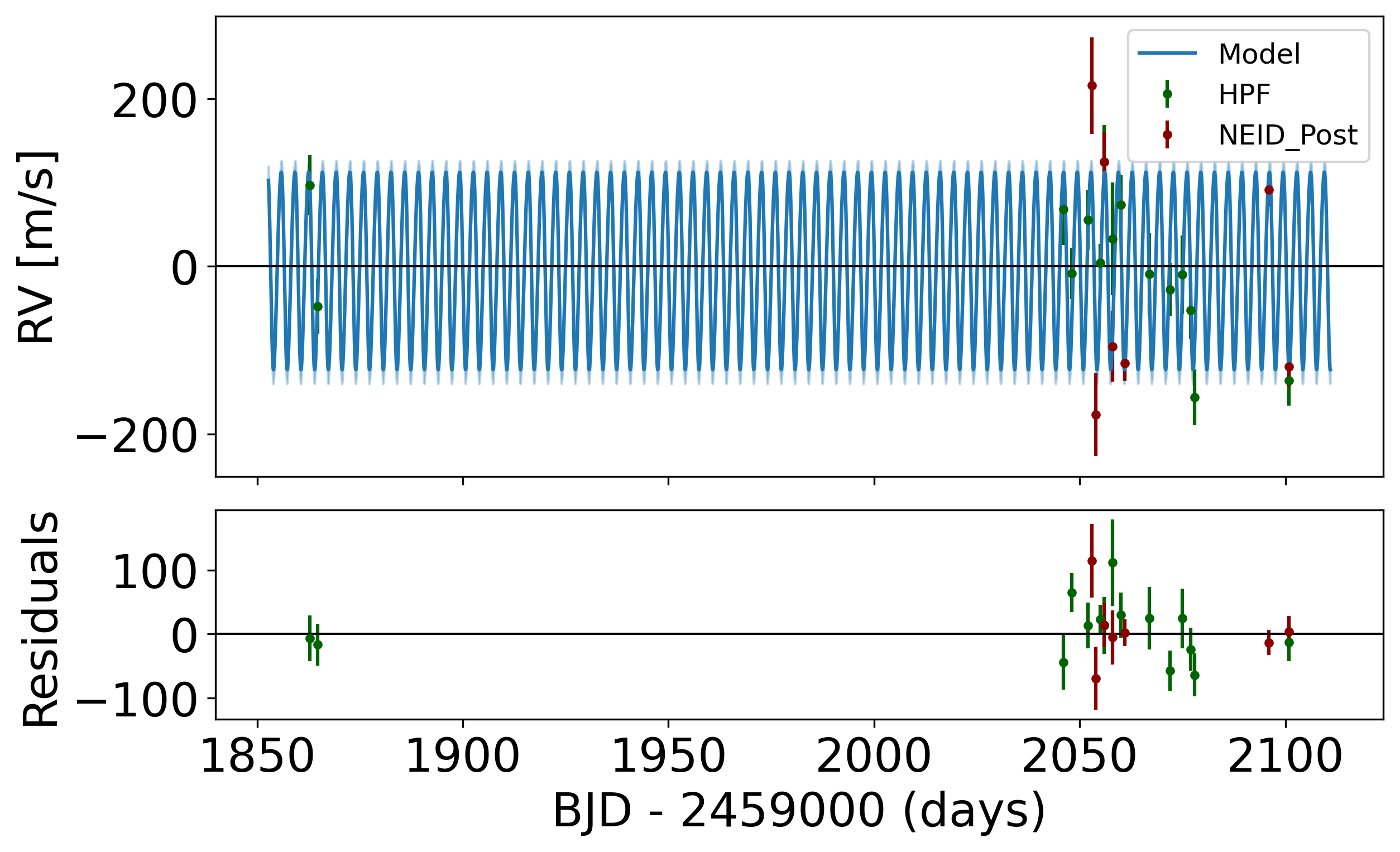}
    \hfill
    \includegraphics[width=0.48\textwidth]{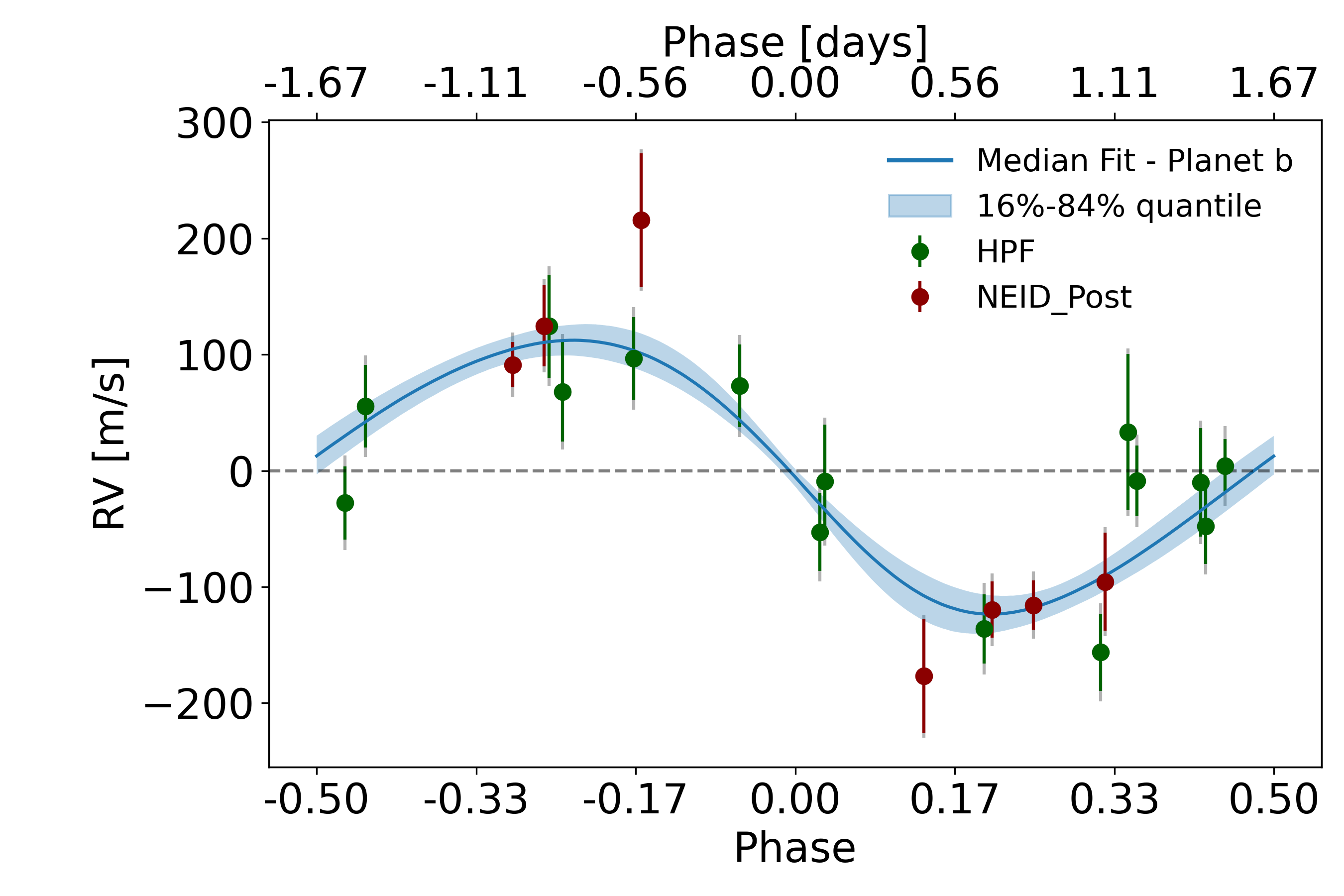}
    \caption{
    Radial velocity observations of \textbf{TOI-7393}.
    \emph{Left:} Combined HPF (Green) and NEID post-break (Maroon) RV time-series with the best-fit Keplerian model.
    \emph{Right:} Phase-folded RVs for TOI-7393\,b.
    }
    \label{fig:rv_7393}
\end{figure*}

\begin{figure*}[t]
    \centering
    \includegraphics[width=0.48\textwidth]{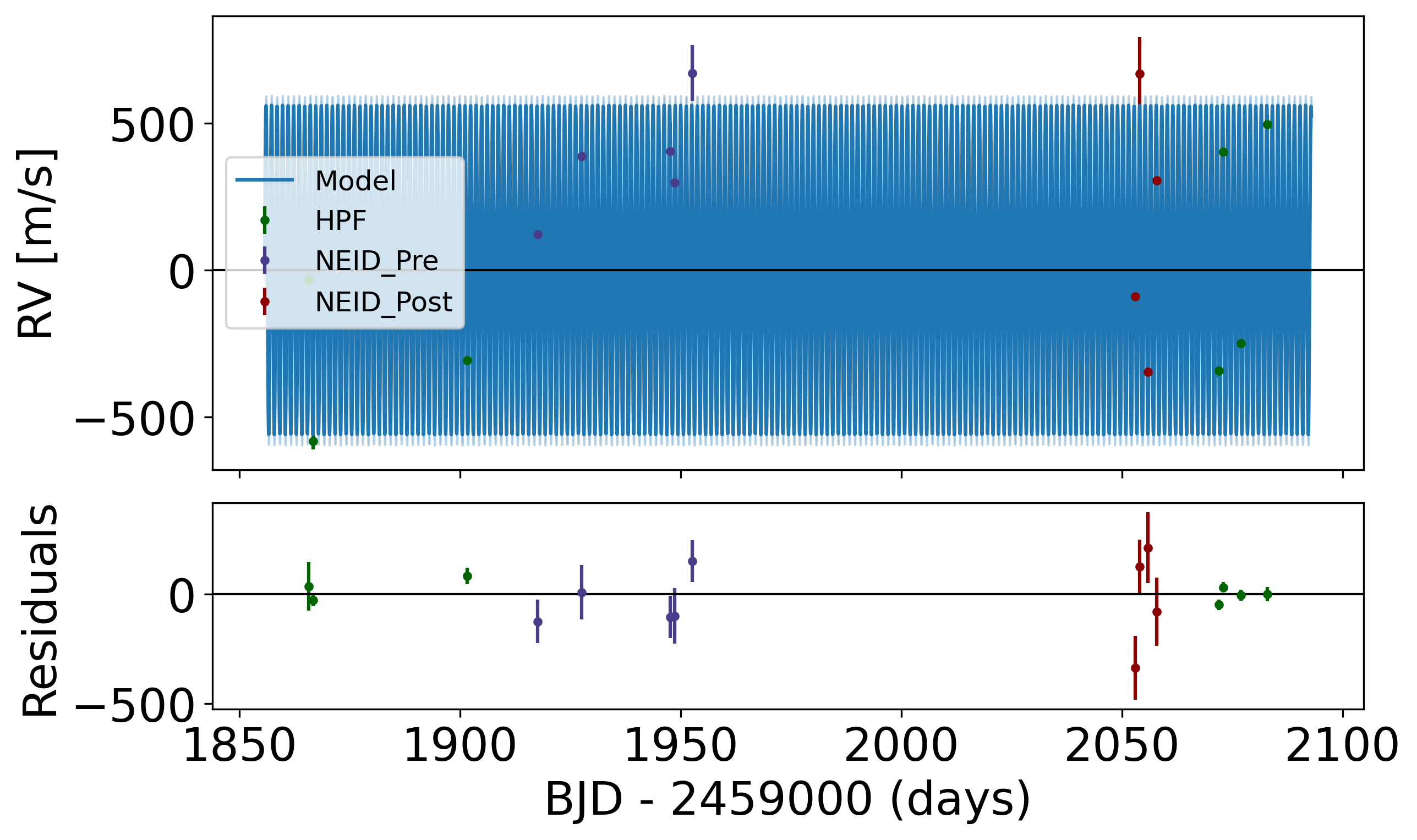}
    \hfill
    \includegraphics[width=0.48\textwidth]{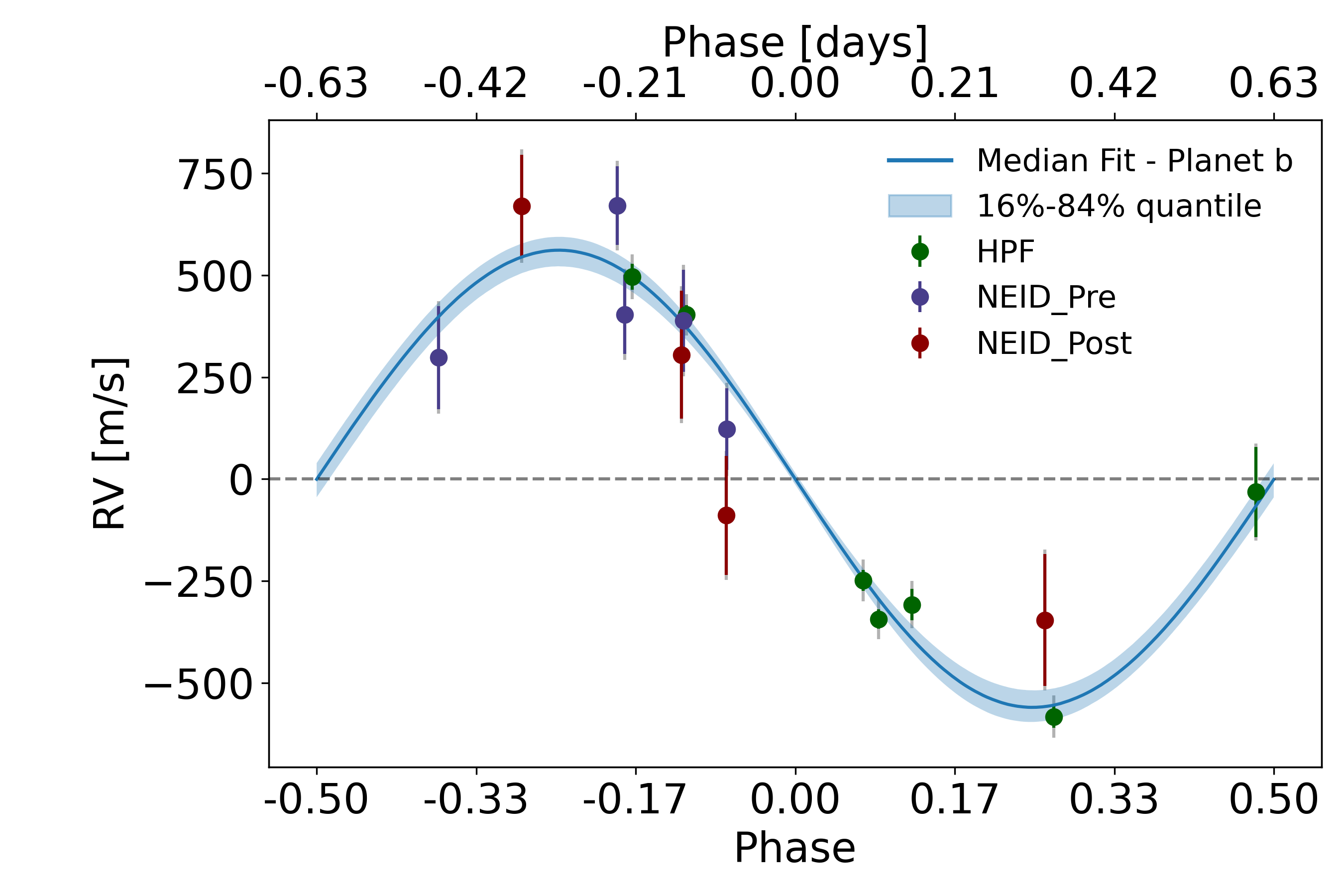}
    \caption{
    Radial velocity observations of \textbf{TOI-7394B}.
    \emph{Left:} Combined HPF (Green), NEID before the RV break on Dec 9th 2025 (Purple), and NEID post-break (Maroon) RV time-series with the best-fit Keplerian model.
    \emph{Right:} Phase-folded RVs for TOI-7394B\,b.
    }
    \label{fig:rv_7394}
\end{figure*}

For TOI-7393 and TOI-7394B, we obtained RV observations from both HPF and NEID, with each target monitored over a baseline of approximately six months with both instruments combined (Figures~\ref{fig:rv_7393} and~\ref{fig:rv_7394}). TOI-7393 was observed with HPF on 14 epochs and with NEID on 5 epochs, all obtained after the NEID RV zero-point change of 2025 December 9\footnote{\url{https://neid.ipac.caltech.edu/docs/NEID-DRP/rveras.html}}. TOI-7394B was observed with HPF on 7 epochs and with NEID on 9 epochs, consisting of 5 measurements acquired prior to the 2025 December NEID RV break and 4 obtained afterward. The pre- and post-break NEID observations were treated as independent instrumental datasets in the RV modeling to account for the potential zero-point offset. All NEID observations were a single exposure of 1800~s. 

All HPF and NEID observations were reduced using their respective standard data reduction pipelines, including bias subtraction, flat-fielding, spectral extraction, and wavelength calibration. The simultaneous wavelength reference was disabled for these M-dwarf targets to avoid contaminating the stellar spectra with calibration light. Instrumental drifts were instead monitored and corrected using regular bracketing calibration exposures, which provide sufficient wavelength stability at the precision level relevant to these observations \citep{metcalf2019, stefansson2020}.

HPF RVs were extracted using a template-matching approach based on the \texttt{SERVAL} pipeline \citep{zechmeister2018}, adapted for HPF spectra \citep{hpfserval}. This method constructs a high SNR stellar template from the observations themselves and determines RV shifts by minimizing residuals between each spectrum and the template across multiple spectral orders.

For targets with NEID observations, two extraction approaches were employed. All spectra were first processed with the NEID Data Reduction Pipeline (DRP)\footnote{\url{https://neid.ipac.caltech.edu/docs/NEID-DRP}}, version~1.5.2. We then used the \texttt{SERVAL} pipeline adapted for NEID spectra \citep{stefansson2022} to compute RVs for TOI-7394B using the 1-D extracted spectra from the DRP. In contrast, RVs for TOI-7393 were obtained directly from the DRP, which calculates velocities using the cross-correlation function (CCF) method. For this target, the CCF mask-based approach yielded higher signal-to-noise RV measurements and reduced scatter compared to template-matching methods, and was therefore adopted for the final analysis.


Prior to RV extraction, spectra were subjected to quality-control cuts to remove observations obtained under poor conditions or with insufficient signal. Spectra were rejected if they exhibited low SNR or anomalously large formal RV uncertainties relative to the median of each dataset, indicative of poor observing conditions or reduction failures. For TOI-7393, two NEID observations were excluded, while for TOI-7394B, one NEID spectrum and one HPF spectrum were removed from the final analysis.

Barycentric velocity corrections were applied using \texttt{barycorrpy}. RV uncertainties reported by the respective extraction pipelines were adopted as the formal measurement errors: for HPF spectra and NEID spectra processed with \texttt{SERVAL}, we use the pipeline-provided uncertainties, while for TOI-7393 we adopt the formal uncertainties from the NEID DRP cross-correlation function (CCF) reduction. These uncertainties reflect photon noise and internal pipeline error propagation.

\subsection{High-Resolution Imaging} \label{sec:imaging}
To assess the presence of unresolved nearby sources that could dilute the transit signals or mimic planetary transits, we obtained high-resolution imaging of all four of our targets using the NN-Explore Exoplanet Stellar Speckle Imager \citep[NESSI;][]{Scott2018} on the WIYN 3.5\,m telescope at Kitt Peak National Observatory.

Data for all of our targets were collected as sequences of 40 ms short-exposure frames (9 sets of 1000 frames) on the night of 2025 October 1 using the SDSS $z^\prime$ and $r^\prime$ filters (Program 2025B-103024; PI: Kanodia). The speckle images were reconstructed as in \citet{Howell2011}. 

\begin{figure*}
    \centering
    \includegraphics[width=0.48\textwidth]{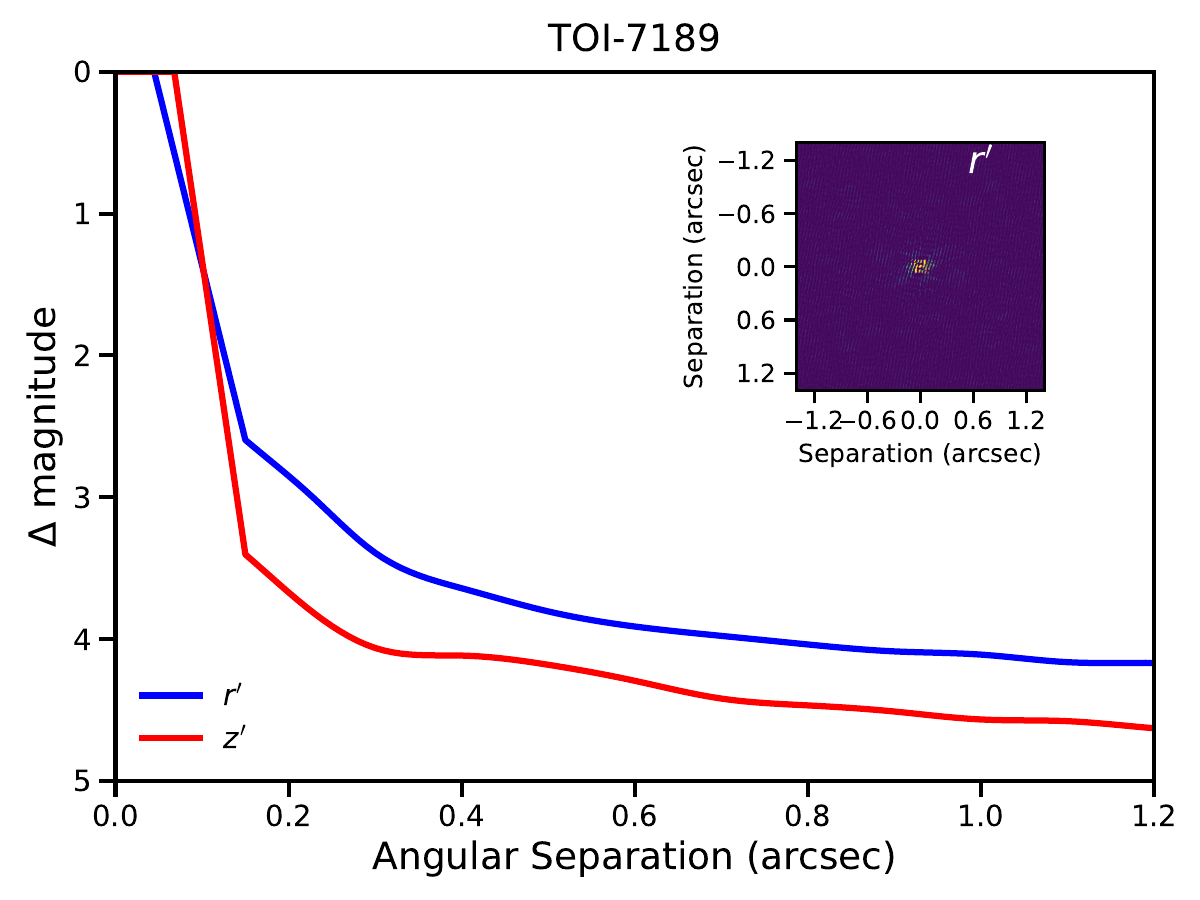}
    \hfill
    \includegraphics[width=0.48\textwidth]{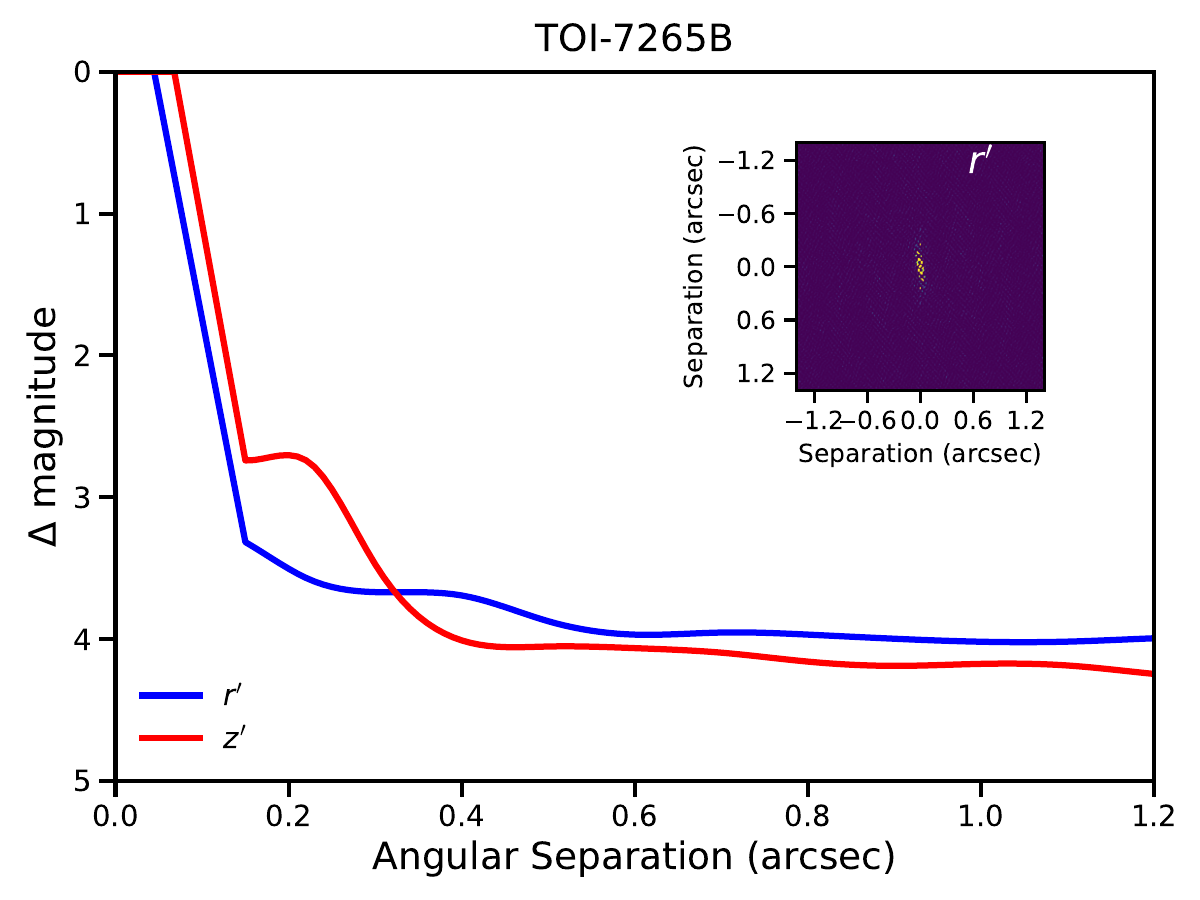}
    \vspace{0.5cm}
    \includegraphics[width=0.48\textwidth]{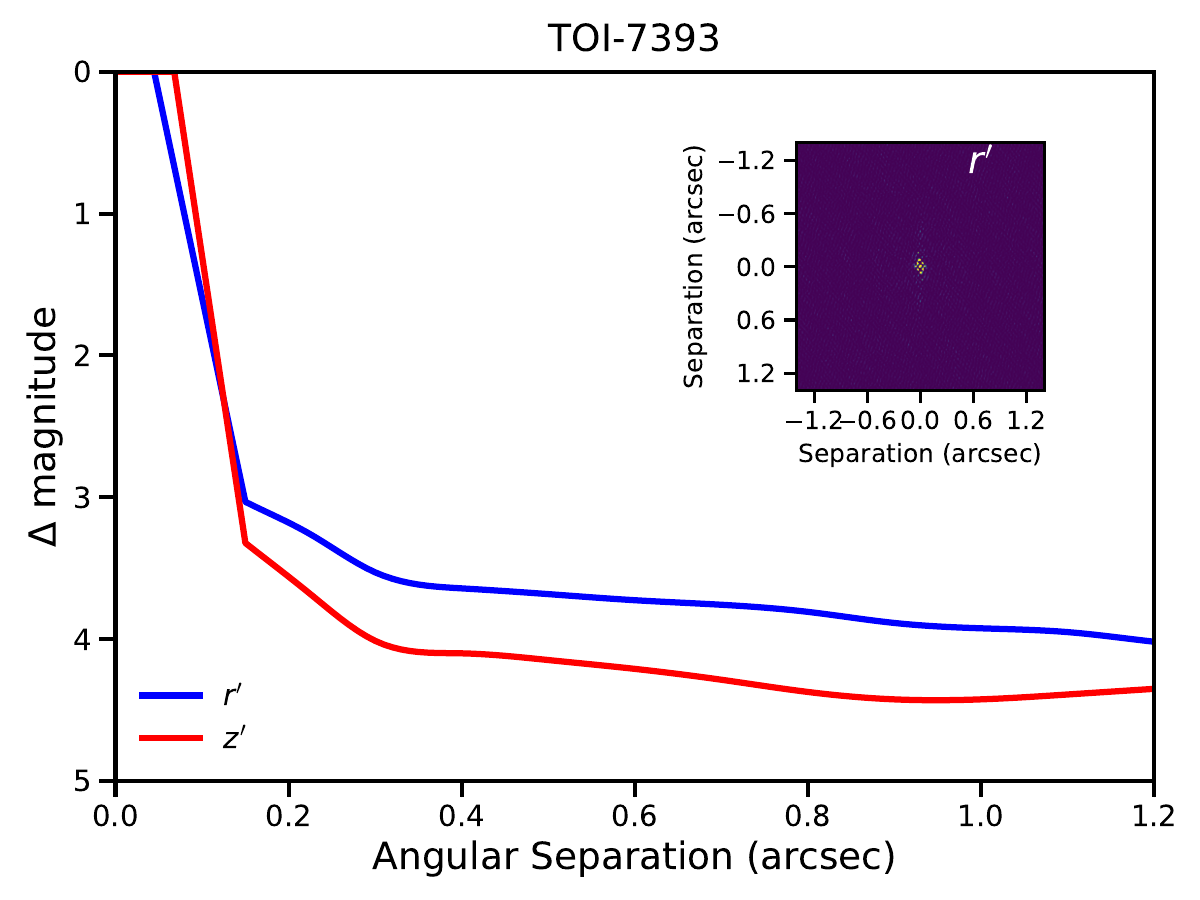}
    \hfill
    \includegraphics[width=0.48\textwidth]{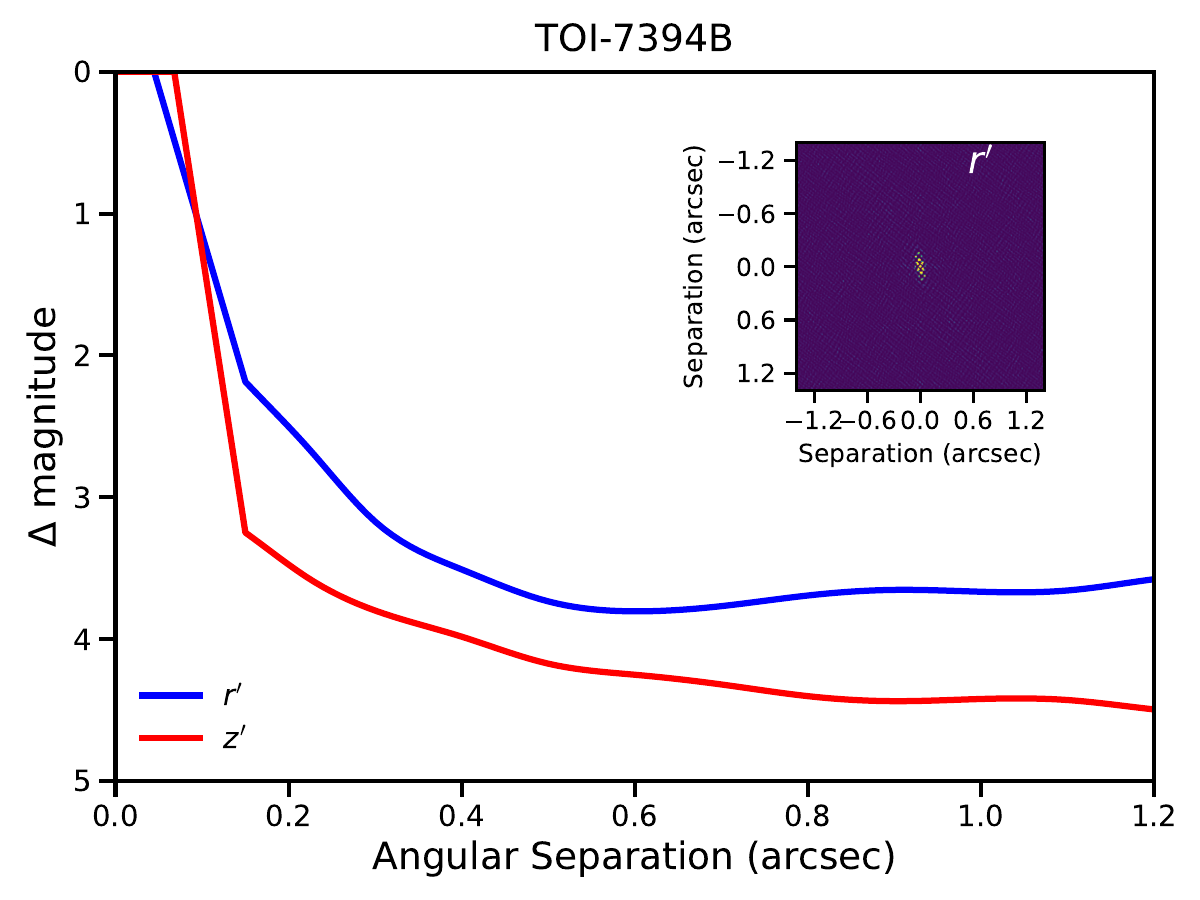}
    \caption{
    High-resolution speckle imaging contrast curves for TOI-7189, TOI-7265B, TOI-7393, and TOI-7394B (top left to bottom right).
    The contrast curves show the $5\sigma$ detection limits as a function of angular separation for the two NESSI filters ($r^\prime$ and $z^\prime$), as indicated in the legend.
    The NESSI observations are sensitive to sources within $1.2\arcsec$.
    No additional sources are detected within these limits for any of the systems, ruling out unresolved stellar companions capable of significantly contaminating the observed transit signals.
    }
    \label{fig:speckle}
\end{figure*}

We note that these observations were acquired during a night affected by known instrumental aberrations in the WIYN optical system, which resulted in increased background structure and periodic artifacts in the reconstructed images and contrast curves. These effects are understood to arise from imperfect point-spread-function matching between science targets and calibration stars and are not astrophysical in origin. Despite these limitations, no nearby sources are detected around either target down to the achieved contrast limits shown in \autoref{fig:speckle}. Additionally, we note the bump in the contrast curve for TOI-7265B originates from the aforementioned instrumental aberrations.

At a separation of $0.5\arcsec$, we rule out companions brighter than $\Delta r^\prime = 3.8$~mag and $\Delta z^\prime = 4.2$~mag for TOI-7189, $\Delta r^\prime = 3.9$~mag and $\Delta z^\prime = 4.1$~mag for TOI-7265B, $\Delta r^\prime = 3.7$~mag and $\Delta z^\prime = 4.1$~mag for TOI-7393 and $\Delta r^\prime = 3.7$~mag and $\Delta z^\prime = 4.2$~mag for TOI-7394B, with increasingly stringent limits at larger separations. These results exclude nearby on-sky stellar companions brighter than the achieved contrast limits that could dilute the transit depths or serve as alternative hosts for the observed signals. 

While speckle imaging alone cannot distinguish between bound companions and background sources, \textit{Gaia} DR3 provides additional constraints on close companions through both resolved sources and astrometric quality indicators. All four targets exhibit renormalized unit weight errors (RUWE) $<$ 1.4, indicating no strong evidence for unresolved companions in the \textit{Gaia} astrometric solutions \citep{elbadry2025}. 

We note that TOI-7394B and TOI-7265B both have wide nearby stars identified in \textit{Gaia} with consistent proper motions and parallaxes. In both systems, the transiting planet orbits the fainter component (designated ``B''), while the brighter nearby star is referred to as the ``A'' component.

The TOI-7394 system is listed in the \citet{elbadry2021} catalog of comoving stellar pairs, with a companion (TOI-7137; hereafter TOI-7394A) at an angular separation of $\sim$3.8\arcsec. Although the inferred distances differ at the $\sim$pc level, this discrepancy is within the uncertainties of the \textit{Gaia} astrometric solutions, and the pair has a very low chance-alignment probability ($\sim10^{-5}$), supporting a likely physical association.

Similarly, TOI-7265 has a wide ($\sim$18\arcsec) nearby star (TOI-7265A) with comparable astrometry, suggestive of a potential comoving companion. In both systems, the companions are sufficiently separated that they do not contaminate the high-resolution spectroscopic observations (fiber diameters $\sim$1--2\arcsec) and have negligible impact on the measured radial velocities. Ground-based follow-up photometry further confirms that the transit signals originate from the fainter target stars (TOI-7394B and TOI-7265B), as described in Section~\ref{sec:ground}.

Together, the high-resolution imaging and \textit{Gaia}-based constraints rule out unresolved stellar companions capable of producing or significantly diluting the observed transit signals, confirming the identified stars as the hosts of the transiting planets.

\section{Host Star Characterization} \label{sec:stellar}

We characterized all four host stars using a combination of high-resolution spectroscopy, broadband photometry, and \textit{Gaia} DR3 astrometry. Spectroscopic parameters were derived from near-infrared spectra obtained with HPF, while global stellar properties were determined through spectral energy distribution (SED) fitting and isochrone modeling using \texttt{EXOFASTv2} \citep{eastman2019}.

\subsection{Spectroscopic Analysis with HPF} \label{sec:spec}

We analyzed the HPF spectra using the \texttt{HPF-SpecMatch}\footnote{\url{https://github.com/gummiks/hpfspecmatch}} framework to derive effective temperature ($T_{\mathrm{eff}}$), surface gravity ($\log g_\star$), metallicity ([Fe/H]), and projected rotational velocity \citep{stefansson2020}. \texttt{HPF-SpecMatch} compares the observed spectra with an empirical library of stars with well-determined parameters, using a linear combination of the five best-matching stars to robustly infer stellar parameters \citep[see][]{Yee2017}. The HPF spectral library used in this work included 100 stars that spanned $2700\mathrm{K} \le T_{\mathrm{eff}} \le 4500~\mathrm{K}$, $4.63<\log g_\star < 5.26$, and $-0.49 < \mathrm{[Fe/H]} < 0.53$.

The resulting spectroscopic parameters are summarized in Table~\ref{tab:stellar_combined}. While labeled as [Fe/H], these metallicities are tied to empirical calibrations for M dwarfs (e.g., \citealt{mann2013, mann2014}) and should be interpreted as a proxy for overall heavy-element abundance ([M/H]-like), rather than a strictly line-by-line iron abundance.

All four targets lie within the parameter space spanned by the \texttt{HPF-SpecMatch} spectral library in $T_{\mathrm{eff}}$, $\log g_\star$, and [Fe/H], and are therefore not extrapolated beyond the bounds of the empirical grid. TOI-7393 lies toward the metal-poor edge of the library, but remains well within the calibrated range. Additionally, the library contains a substantial number of stars with effective temperatures comparable to our early-M dwarf targets, ensuring robust spectral matches.

\begin{figure*}
    \centering
    \includegraphics[width=\textwidth]{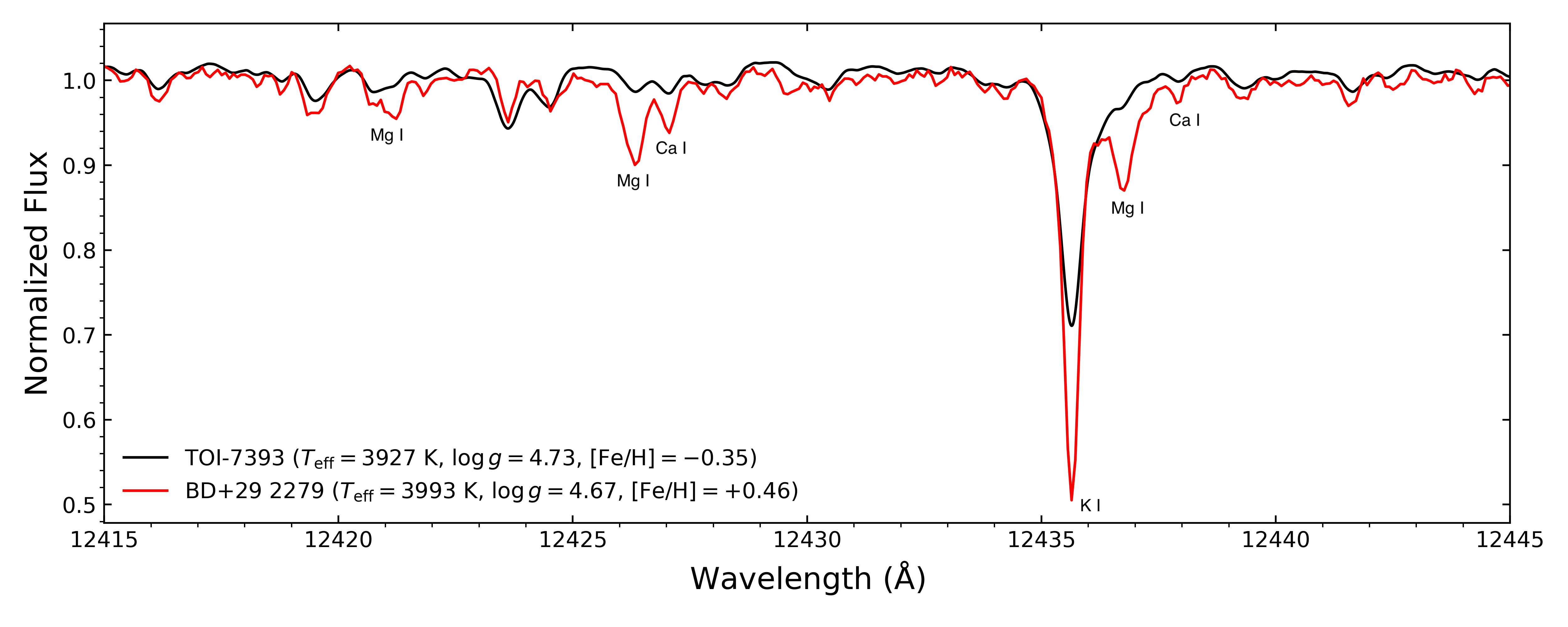}
\caption{
HPF spectrum of TOI-7393 (black) compared to the metal-rich reference star BD+29~2279 (red) over a 30~\AA\ region near the K~I line at $\sim$12435.7~\AA. The two stars have similar effective temperatures and surface gravities, but TOI-7393 ($\mathrm{[Fe/H]}=-0.35$) exhibits systematically weaker absorption features than BD+29~2279 ($\mathrm{[Fe/H]}=+0.46$), consistent with its lower metallicity. Various atomic species are shown for reference, from Vienna Atomic Line Database (VALD) \citep{ryabchikova2015}.
}
    \label{fig:toi7393_spec}
\end{figure*}

Three of the host stars (TOI-7189, TOI-7265B, and TOI-7394B) are metal-rich early-M dwarfs. In contrast, TOI-7393 displays a significantly sub-solar metallicity ($\mathrm{[Fe/H]} \approx -0.35$), making it among the more metal-poor giant-planet hosts currently identified in the GEMS sample. This difference is also evident directly in the HPF spectra (Figure~\ref{fig:toi7393_spec}), where TOI-7393 exhibits systematically weaker absorption features compared to a metal-rich reference star of similar temperature and surface gravity. Its Galactic kinematics place it in the lower end of the thin/thick-disk transition regime at $TD/D \sim 0.6$, consistent with an older stellar population and in line with its sub-solar metallicity \citep{haywood2008}. This is discussed further in Section~\ref{sec:uvw}.

As noted by \cite{passeger2022} and \cite{kanodia2024a}, the complexities of M-dwarf spectra limit the accuracy of \texttt{HPF-SpecMatch} for metallicity determinations, and we advise caution in interpreting the [Fe/H] values beyond qualitative trends. Nevertheless, the relative differences between the systems are robust: TOI-7189, TOI-7265B, and TOI-7394B appear metal-rich, consistent with the well-established correlation between stellar metallicity and giant planet occurrence \citep{gan2025, han2024}, whereas TOI-7393 provides a notable counterexample demonstrating that gas-rich planets can form even around comparatively metal-poor M dwarfs, albeit likely with lower efficiency.

As an independent check, we also estimated the stellar metallicities using the photometric calibrations of \citet{duque2023} implemented in the \texttt{metamorphosis} tool \citep{duque2022}, which combines broadband photometry with \textit{Gaia} astrometry. Adopting the relations based on \textit{Gaia} $G$ and 2MASS near-infrared magnitudes (excluding $G_{\rm BP}$/$G_{\rm RP}$ and \textit{WISE} bands to minimize potential systematics for faint M dwarfs), we obtain photometric [Fe/H] values that are consistent within uncertainties with those derived from \texttt{HPF-SpecMatch}. While the true metallicities remain subject to the intrinsic limitations of both spectroscopic modeling and photometric calibrations, the agreement between these independent methods supports the robustness of the relative metallicity ordering among the four systems.

The measured low projected rotational velocities for all four stars are consistent with relatively slow rotators and broadly agree with the modest activity levels inferred from spectroscopic diagnostics \citep{west2015}, a conclusion further supported by the absence of detectable photometric rotation signals (Section~\ref{sec:rotation}).

\subsection{Photometric Rotation Constraints} \label{sec:rotation}

We searched for photometric rotation signals using the \texttt{TGLC} light curves, and publicly available Zwicky Transient Facility (\textit{ZTF}; \citet{bellm2014}) light curves for all four targets, from Data Release 24. In both datasets, we computed Generalized Lomb--Scargle periodograms \citep{zechmeister2009} over a period range of 0.5--100~days, which encompasses the sensitivity window set by the observing baselines and cadences.

In the \textit{TESS} photometry, no significant periodic signals were detected for any target, consistent with low-amplitude variability below the detection threshold. The \textit{ZTF} periodograms are dominated by strong peaks near 1~day and its harmonics, consistent with known diurnal sampling aliases in ground-based time-series data. No additional peaks exceed a false-alarm probability (FAP) threshold of 1\% across the explored period range.

We therefore find no statistically significant evidence for rotational modulation in either dataset. This non-detection is broadly consistent with the projected rotational velocities inferred from the HPF spectroscopy. For TOI-7189, TOI-7265B, and TOI-7394B, the measured line broadening is below the $\sim$2 km\,s$^{-1}$ threshold at which rotational broadening can be robustly distinguished at HPF's spectral resolution. TOI-7393 yields a projected rotational velocity of $v\sin i_\star = 4.24 \pm 1.97$ km\,s$^{-1}$, although the measurement remains only marginally above this practical detection limit. Together, these measurements indicate that none of the host stars are rapid rotators. The absence of detectable periodic variability further suggests that the stellar rotation periods likely exceed the sensitivity limits of the available photometry.

\subsection{SED and Isochrone Modeling} \label{sec:sed}
We derived global stellar parameters by modeling the spectral energy distribution (SED) with \texttt{EXOFASTv2}\footnote{\url{https://github.com/jdeast/EXOFASTv2}} \citep{eastman2019}. 
Inputs included \textit{Gaia} DR3 parallaxes, broadband photometry from PanSTARRS1 DR2 \citep{Chambers2016}, 2MASS \citep{skrutskie2006}, and \textit{WISE}\citep{wright2010}, and Gaussian priors on $T_{\rm eff}$, $\log g_\star$, and [Fe/H] derived from the HPF spectroscopic analysis, as listed in Table~\ref{tab:stellar_combined}.

We further constrained the interstellar extinction using three-dimensional dust maps \citep{green2019}, allowing it to vary within a narrow range appropriate for stars at the distances of our targets, and adopted an $R=3.1$ reddening law from \citet{Fitzpatrick1999}.



The SED and isochrone fits yield stellar masses, radii, luminosities, and mean densities consistent with mid-M dwarf evolutionary models \citep{dotter2008, choi2016}. The derived stellar parameters are listed in Table~\ref{tab:stellar_combined}. We find no evidence for significant excess emission or unresolved companions that would bias the photometric modeling. Stellar ages are not well constrained by SED and isochrone fitting for main-sequence M dwarfs, as their evolutionary tracks change very slowly; we therefore do not place strong constraints on the ages of these systems.


\subsection{Consistency Checks and Multiplicity Constraints}

As an internal consistency check, we compared the stellar densities inferred from the SED and isochrone analysis \citep{eastman2019} to the mean stellar densities derived from the transit light curves under the assumption of circular orbits. For all four of our targets, the density estimates agree within $<2\sigma$, supporting the adopted stellar parameters and indicating that unmodeled eccentricity or photometric dilution is unlikely to significantly impact our results. In addition, high-resolution imaging and contrast-curve constraints (Section~\ref{sec:imaging}) rule out nearby stellar companions capable of producing the observed transit signals or significantly contaminating the \textit{TESS} photometry.

\begin{sidewaystable*}[ht!]
\centering
\caption{
Stellar and ancillary properties of TOI-7189, TOI-7265B, TOI-7393 and TOI-7394B.
Astrometry and photometry are primarily from Gaia Data Release 3 and complementary ground-based surveys. Spectroscopic parameters are derived from HPF spectroscopy, while model-dependent stellar properties are obtained from SED and isochrone fitting. The uncertainties reported for the HPF SpecMatch spectroscopic parameters correspond to the cross-validation uncertainties of the empirical M-dwarf spectral library adopted by the method \citep{Yee2017, stefansson2020}; because the same underlying library calibration is used for all targets, several parameter uncertainties are identical across the sample.
}
\label{tab:stellar_combined}
\scriptsize
\begin{tabular*}{\textheight}{@{\extracolsep{\fill}}llcccc l}
\hline\hline
Parameter & Description & TOI-7189 & TOI-7265B & TOI-7393 & TOI-7394B & Reference \\
\hline
\multicolumn{7}{l}{\textbf{Main Identifiers:}} \\
TOI & TESS Object of Interest & 7189 & 7265B & 7393 & 7394B & TESS Mission \\
TIC & TESS Input Catalog & 18885659 & 275717567 & 165598669 & 161687211 & TESS Input Catalog \\
Gaia DR3 & Gaia Data Release 3 source ID & 4276149124440902784 & 2164125035679377664 & 1600846431341704704 & 1602132340256374528 & Gaia DR3 \\
\hline
\multicolumn{7}{l}{\textbf{Equatorial Coordinates, Proper Motion, \& Parallax:}} \\
$\alpha$ & Right Ascension (ICRS epoch 2015.5) & 18:12:09 & 21:14:48.96 & 15:09:26.36 & 15:44:13.25 & Gaia DR3 \\
$\delta$ & Declination (ICRS epoch 2015.5) & $+$02:11:17.74 & $+$45:55:33.27 & $+$55:58:51.8 & $+$58:33:09.94 & Gaia DR3 \\
$\mu_\alpha$ & Proper motion in RA (mas yr$^{-1}$) & $3.722 \pm 0.073$ & $30.024 \pm 0.088$ & $-40.156 \pm 0.032$ & $-25.997 \pm 0.053$ & Gaia DR3 \\
$\mu_\delta$ & Proper motion in Dec (mas yr$^{-1}$) & $-22.469 \pm 0.077$ & $-28.743 \pm 0.089$ & $-76.945 \pm 0.030$ & $0.214 \pm 0.056$ & Gaia DR3 \\
$\varpi$ & Parallax (mas) & $5.21 \pm 0.03$ & $6.96 \pm 0.03$ & $5.27 \pm 0.01$ & $5.18 \pm 0.02$ & Gaia DR3 \\
$d$ & Distance (pc) & $192.09 \pm 1.60$ & $143.97 \pm 1.05$ & $189.04 \pm 0.65$ & $189.99 \pm 0.96$ & TESS Input Catalog \\
\hline
\multicolumn{7}{l}{\textbf{Optical \& Near-Infrared Magnitudes:}} \\
$G$ & Gaia $G$ magnitude & $14.787 \pm 0.002$ & $15.253 \pm 0.002$ & $14.287 \pm 0.003$ & $14.741 \pm 0.001$ & Gaia DR3 \\
$G_{\rm BP}$ & Gaia $G_{\rm BP}$ magnitude & $15.604 \pm 0.010$ & $16.179 \pm 0.013$ & $15.100 \pm 0.003$ & $15.772 \pm 0.003$ & Gaia DR3 \\
$G_{\rm RP}$ & Gaia $G_{\rm RP}$ magnitude & $14.239 \pm 0.005$ & $14.682 \pm 0.007$ & $13.397 \pm 0.004$ & $13.713 \pm 0.001$ & Gaia DR3 \\
$g$ & PS1 $g$ magnitude & $16.399 \pm 0.033$ & $16.972 \pm 0.003$ & $15.436 \pm 0.004$ & $16.204 \pm 0.009$ & PS1 DR2 \\
$r$ & PS1 $r$ magnitude & $15.183 \pm 0.005$ & $15.813 \pm 0.003$ & $14.342 \pm 0.005$ & $14.973 \pm 0.006$ & PS1 DR2 \\
$i$ & PS1 $i$ magnitude & $14.278 \pm 0.006$ & $14.551 \pm 0.002$ & $13.794 \pm 0.004$ & $14.142 \pm 0.001$ & PS1 DR2 \\
$z$ & PS1 $z$ magnitude & $13.856 \pm 0.0017$ & $13.994 \pm 0.004$ & $13.547 \pm 0.006$ & $13.781 \pm 0.001$ & PS1 DR2 \\
$y$ & PS1 $y$ magnitude & $13.651 \pm 0.005$ & $13.676 \pm 0.005$ & $13.376 \pm 0.004$ & $13.559 \pm 0.002$ & PS1 DR2 \\
$J$ & 2MASS $J$ magnitude & $12.309 \pm 0.026$ & $12.419 \pm 0.028$ & $12.304 \pm 0.025$ & $12.377 \pm 0.026$ & 2MASS \\
$H$ & 2MASS $H$ magnitude & $11.659 \pm 0.023$ & $11.753 \pm 0.024$ & $11.667 \pm 0.024$ & $11.726 \pm 0.023$ & 2MASS \\
$K_s$ & 2MASS $K_s$ magnitude & $11.430 \pm 0.020$ & $11.508 \pm 0.023$ & $11.477 \pm 0.022$ & $11.492 \pm 0.023$ & 2MASS \\
$W1$ & WISE $W1$ magnitude & $11.261 \pm 0.022$ & $11.357 \pm 0.023$ & $11.417 \pm 0.023$ & $10.432 \pm 0.022$ & WISE \\
$W2$ & WISE $W2$ magnitude & $11.305 \pm 0.020$ & $11.361 \pm 0.022$ & $11.391 \pm 0.021$ & $10.499 \pm 0.02$ & WISE \\
\hline
\multicolumn{7}{l}{\textbf{Spectroscopic Parameters:}} \\
$T_{\rm eff}$ & Effective temperature (K) & $3775\pm59$ & $3503 \pm 59$ & $3927 \pm 59$ & $3717\pm59$ & This work \\
{[Fe/H]} & Metallicity (dex) & $+0.47 \pm 0.16$ & $+0.35 \pm 0.16$ & $-0.35\pm0.16$ & $+0.50\pm0.16$ & This work \\
$\log g_\star$ & Surface gravity (cgs) & $4.67 \pm 0.04$ & $4.77 \pm 0.04$ & $4.73 \pm 0.04$ & $4.66\pm0.04$ & This work \\
$v\sin i_\star$ & Projected rotation (km s$^{-1}$) & $< 2$ & $< 2$ & $4.24\pm1.97$ & $< 2$ & This work \\
\hline
\multicolumn{7}{l}{\textbf{Model-Dependent Stellar Parameters (SED \& Isochrone Fits):}} \\
$M_\star$ & Stellar mass ($M_\odot$) & $0.612^{+0.024}_{-0.023}$ & $0.521 \pm 0.023$ & $0.600 \pm 0.021$ & $0.624 \pm 0.022$ & This work \\
$R_\star$ & Stellar radius ($R_\odot$) & $0.591^{+0.015}_{-0.014}$ & $0.497 \pm 0.013$ & $0.582 \pm 0.013$ & $0.609 \pm 0.015$ & This work \\
$L_\star$ & Stellar luminosity ($L_\odot$) & $0.065 \pm 0.003$ & $0.034 \pm 0.001$ & $0.078 \pm 0.002$ & $0.068 \pm 0.002$ & This work \\
Spectral Type & Classification & M1 dwarf & M2 dwarf & M0 dwarf & M1 dwarf & This work \\
\hline
\multicolumn{7}{l}{\textbf{Systemic Velocity \& Galactic Kinematics:}} \\
$\gamma$ & Systemic RV (km s$^{-1}$) 
& 0.04  $\pm$ 0.13
& -62.03 $\pm$ 0.16
& -5.26 $\pm$ 1.05 
& -17.68 $\pm$ 1.00
& This work \\

$U$ & Galactic velocity toward GC (km s$^{-1}$)
& $10.26$ 
& $-1.57$ 
& $37.78$ 
& $-10.19$ 
& This work \\

$V$ & Galactic rotation velocity (km s$^{-1}$)
& $-13.29$ 
& $-62.96$ 
& $-56.40$ 
& $-27.47$ 
& This work \\

$W$ & Vertical Galactic velocity (km s$^{-1}$)
& $-12.17$ 
& $-26.18$ 
& $38.32$ 
& $1.34$ 
& This work \\
\hline
\hline
\end{tabular*}
\end{sidewaystable*}

\subsection{Galactic Kinematics (UVW)} \label{sec:uvw}

We computed Galactic space velocities for TOI-7189, TOI-7265B, TOI-7393, and TOI-7394B using \textit{Gaia} DR3 astrometry in combination with the systemic radial velocities derived from our global modeling (Section~\ref{sec:modeling}). 

The calculations were performed using the \texttt{pyasl.gal\_uvw} routine from the \texttt{PyAstronomy} package \citep{pya}, which implements the standard transformation from equatorial coordinates, proper motions, parallax, and radial velocity into a right-handed Galactic Cartesian velocity system. 

We adopt the convention of \citet{johnson1987}, with $U$ positive toward the Galactic center, $V$ positive in the direction of Galactic rotation, and $W$ positive toward the north Galactic pole. The systemic velocities and derived Galactic space motions ($U,V,W$) for all four systems are reported in Table~\ref{tab:stellar_combined}.

To classify the Galactic population membership of the host stars, we computed thin-disk, thick-disk, and halo probabilities following the kinematic prescription of \citet{bensby2014}, in which each population is modeled as a Gaussian velocity ellipsoid with characteristic dispersions, asymmetric drift, and local normalization. Using the derived $(U,V,W)$ velocities, we evaluated the relative probabilities of thick-disk and halo membership with respect to the thin disk, denoted as $TD/D$ and $H/D$, respectively.

Two systems (TOI-7189 and TOI-7394B) exhibit $TD/D < 0.1$, consistent with strong thin-disk membership. TOI-7265B and TOI-7393 yield moderately elevated values of $TD/D \sim 0.4$ and $0.6$, respectively. Under the \citet{bensby2014} prescription, these values approach the transition regime between the thin and thick disks, although they remain more consistent with thin-disk membership than with the higher $TD/D$ values typically associated with thick-disk populations \citep{bensby2003}. The elevated values are driven primarily by their significant lag in $V$ and enhanced velocity dispersions, particularly in $W$, and highlight the sensitivity of population classification to the adopted kinematic prescription. We note that alternative formalisms (e.g., \citealt{reddy2006}) yield comparable probabilities for thin- and thick-disk membership for these systems. All targets have negligible halo probabilities ($H/D \ll 1$), indicating that none are consistent with halo kinematics.

\subsection{Comparison with Empirical M-dwarf Relations} \label{sec:mann_check}
Some of the stellar parameters adopted in this work were derived using \texttt{EXOFASTv2} with MIST stellar evolution models. However, radii and masses of M dwarfs are known to be challenging to model accurately using theoretical tracks alone \citep[e.g.,][]{kesseli2018}. Because the inferred planetary radii and bulk densities depend directly on the host star properties, we compared our model-derived stellar parameters to empirical M-dwarf relations from \citet{mann2019} as an independent consistency check.

Using the \citet{mann2019} relations, which estimate stellar properties from calibrated relations involving absolute near-infrared magnitudes and metallicity, we obtain stellar masses of
$M_\star = 0.600 \pm 0.015\,M_\odot$ (TOI-7189), $0.491 \pm 0.012\,M_\odot$ (TOI-7265B), $0.588 \pm 0.015\,M_\odot$ (TOI-7393), and $0.587 \pm 0.015\,M_\odot$ (TOI-7394B). These empirically derived values are consistent with the masses inferred from our global modeling within $<1.2\sigma$ for all systems, supporting the robustness of the adopted stellar parameters.

We note that systematic offsets between model-based and empirical M-dwarf radii at the few-percent level could propagate into the inferred planetary radii and densities. However, the agreement between the MIST-based and empirical mass estimates suggests that any such systematics are unlikely to qualitatively affect our conclusions regarding the Saturn-like densities and bulk compositions of the companions. 

\section{Joint Modeling of Photometry and Radial Velocities} \label{sec:modeling}

We jointly modeled the \textit{TESS} photometry, ground-based light curves, and radial velocity measurements for TOI-7265B\,b, TOI-7189\,b, TOI-7393\,b, and TOI-7394B\,b using the \texttt{exoplanet} modeling framework \citep{foremanmackey2021} together with \texttt{pyMC3} \citep{salvatier2016}. \texttt{pyMC3} is a probabilistic programming framework for Bayesian statistical modeling that enables efficient sampling of complex posterior distributions using advanced Markov Chain Monte Carlo (MCMC) algorithms, such as the No-U-Turn Sampler (NUTS) \citep{hoffman2014}. Within this framework, \texttt{exoplanet} provides specialized tools for modeling exoplanet transit and radial velocity signals while leveraging \texttt{pyMC3} to perform posterior inference. This approach enables a fully Bayesian treatment of the transit and RV data, allowing correlated uncertainties and physically motivated priors to be propagated self-consistently through the global fit.

Free parameters in the joint model included the orbital period ($P$), time of inferior conjunction ($T_0$), planet-to-star radius ratio ($R_p/R_*$), scaled semi-major axis ($a/R_*$), impact parameter ($b$), inclination ($i$), eccentricity ($e$), argument of periastron ($\omega$), RV semi-amplitude ($K$), and instrument-specific systemic velocity offsets and jitter terms. Gaussian priors on the stellar mass, radius, and effective temperature were adopted from the SED and isochrone analysis described in Section~\ref{sec:stellar}. Limb-darkening coefficients for each instrument were parameterized using the $(q_1, q_2)$ transformation of \citet{kipping2013} and sampled with uniform priors over the physically allowed unit disk.

Posterior sampling was performed using NUTS. Prior to sampling, the model parameters were optimized to obtain a maximum a posteriori (MAP) solution \citep{bassett2016}, which was used to initialize the MCMC chains. Convergence was assessed using the Gelman–Rubin statistic ($\hat{R}$) and by visually inspecting the MCMC chains. All sampled parameters satisfied $\hat{R} < 1.01$, indicating good convergence. Quantities not directly sampled in the fit were derived from the posterior samples. Both sampled and derived posteriors are reported in Table~\ref{tab:planetparams}, with derived parameters explicitly indicated.

\subsection{Transit Model}

Transit light curves for all four systems were modeled using the analytic formalism of \citet{mandel2002}. The \texttt{TGLC} products provide \textit{TESS} light curves corrected for crowding using Gaia-based information; however, these corrections are not guaranteed to fully capture residual blending given the large \textit{TESS} pixel scale and uncertainties in the contaminating flux estimates. To account for potential under- or over-correction, we included an additional dilution parameter in the transit model for each \textit{TESS} dataset. This parameter does not impose the \texttt{TGLC} correction as a prior, but instead allows the fitted transit depth to adjust for any residual mismatch between the assumed and actual level of contaminating flux.

In contrast, the ground-based light curves, obtained at significantly higher angular resolution and verified to be free of nearby stellar contaminants within the photometric apertures, were modeled assuming no dilution. In this framework, the \textit{TESS} photometry primarily constrains the transit shape and timing, while the ground-based transits constrain the transit depth. The fitted dilution factors for the \textit{TESS} data are generally modest, with values ranging from $\sim$0.6 to 1.3 (Table~\ref{tab:appendix_jitter}), indicating that the \texttt{TGLC} corrections broadly agree with the ground-based measurements within uncertainties.

The transit signals for all four systems are well described by a single, periodic transit model, with consistent depths and durations across all observed \textit{TESS} sectors and ground-based observations and no evidence for significant transit timing variations. TOI-7393\,b and TOI-7394B\,b exhibit high-impact-parameter transit geometries, with impact parameters of $b = 0.75^{+0.05}_{-0.06}$ and $b = 0.74^{+0.03}_{-0.04}$, respectively. TOI-7393\,b is consistent with a grazing or near-grazing configuration, as the posterior distribution allows $b + R_p/R_* > 1$, producing a V-shaped transit morphology. In contrast, TOI-7394B\,b remains consistent with a high-latitude full transit ($b + R_p/R_* < 1$), yielding a rounded, U-shaped profile despite the large impact parameter.

\subsection{Radial Velocity Model}

The RV data were modeled simultaneously with the photometry using Keplerian orbits, allowing for independent RV offsets and jitter terms for each instrument, and the two NEID RV eras for TOI-7394B. Generalized Lomb-Scargle periodograms of the RVs show significant power at the photometric orbital periods for all four systems, with FAP below 1\%, supporting a planetary origin for the observed RV signals. Additional astrophysical and instrumental noise was accounted for through instrument-specific RV jitter terms included in the joint modeling. 

To account for additional astrophysical variability and residual instrumental systematics not captured by the formal RV uncertainties, we include an instrument-specific jitter term in the joint model. This term is added in quadrature to the reported measurement uncertainties within the likelihood function and is inferred simultaneously with the orbital parameters.

The inferred RV semi-amplitudes are significant and consistent with the photometric ephemerides for all four systems, yielding well-constrained planetary masses (Table~\ref{tab:planetparams}). The planets span a factor of $\sim4$ in mass, from Saturn-mass objects ($\sim0.5$--$0.7\,M_{\rm J}$) to a super-Jupiter in TOI-7394B\,b ($\sim2.1\,M_{\rm J}$), highlighting the diversity of giant planets around early-M dwarfs.

The orbits of all four planets are consistent with low eccentricities, as expected for short-period systems that are likely to have undergone tidal circularization. The inferred separations ($a/R_\star$ = 7-17) correspond to eccentricity damping timescales of $\sim10^{6}$-$10^{8}$ yr \citep{jackson2008}, which are short compared to the typical main-sequence lifetimes of M dwarfs ($\sim10^{11}$-$10^{12}$ yr; \citealt{engle2023}). The equilibrium temperatures of the planets are $T_{\rm eq} = 738 \pm 32$~K for TOI-7189\,b, $571 \pm 27$~K for TOI-7265B\,b, $747 \pm 32$~K for TOI-7393\,b, and $986 \pm 42$~K for TOI-7394B\,b. These temperatures place the planets in the warm giant-planet regime around M dwarfs, where relatively few well-characterized transiting systems with precise mass and radius measurements currently exist \citep{kanodia2024a, kanodia2024b}. 

We also tested for the presence of long-term accelerations in the RV data that could indicate additional massive companions in wider orbits. For each system, we included both linear and quadratic RV trend terms in the joint modeling and evaluated their statistical significance. In all cases, the inferred trend coefficients were consistent with zero to within $1\sigma$, and their inclusion did not improve the Bayesian Information Criterion (BIC). We therefore adopt the simpler no-trend model as the preferred solution.

As a robustness check, we verified that allowing for linear or quadratic trends does not significantly affect the inferred planetary parameters: the derived masses remain consistent to well within $1\sigma$ when trend terms are included.

\begin{sidewaystable*}[ht!]
\centering
\caption{
Derived planetary parameters for TOI-7189\,b, TOI-7265B\,b, TOI-7393\,b, and TOI-7394B\,b derived from joint fits to \textit{TESS} photometry, ground-based transit observations, and HPF radial velocities.
}
\label{tab:planetparams}
\footnotesize
\begin{tabular*}{\textwidth}{@{\extracolsep{\fill}}llcccc}
\hline\hline
Parameter & Description & TOI-7189\,b & TOI-7265B\,b & TOI-7393\,b & TOI-7394B\,b \\
\hline
\multicolumn{6}{c}{\textbf{Orbital Parameters}} \\
\hline
$P$ & Orbital period (days) &
$2.89055^{+0.00000612}_{-0.00000610}$ &
$4.17232^{+0.00000256}_{-0.00000257}$ &
$3.33708^{+0.00000095}_{-0.00000094}$ &
$1.252995^{+0.00000031}_{-0.00000031}$ \\

$e$ & Eccentricity &
$0.0601^{+0.0615}_{-0.0425}$ &
$0.0372^{+0.0421}_{-0.0258}$ &
$0.1198^{+0.0651}_{-0.0627}$ &
$0.0271^{+0.0305}_{-0.0193}$ \\

$\omega$ & Argument of periastron (rad) &
$-0.177^{+1.56}_{-1.19}$ &
$-0.001^{+2.36}_{-2.39}$ &
$2.00^{+0.571}_{-0.759}$ &
$0.435^{+1.79}_{-2.44}$ \\

$K$ & RV semi-amplitude (m\,s$^{-1}$) &
$104.7^{+18.8}_{-18.5}$ &
$143.5^{+18.1}_{-18.4}$ &
$119.7^{+13.0}_{-12.9}$ &
$562.2^{+30.3}_{-36.8}$ \\

\hline
\multicolumn{6}{c}{\textbf{Transit Parameters}} \\
\hline
$T_C$ & Transit midpoint ($BJD_{TDB}$) &
$2460503.94932^{+0.000536}_{-0.000541}$ &
$2460576.09300^{+0.000367}_{-0.000348}$ &
$2460369.40819^{+0.000303}_{-0.000302}$ &
$2460556.90137^{+0.000243}_{-0.000241}$ \\

$R_p/R_\star$ & Radius ratio &
$0.1731^{+0.00544}_{-0.00550}$ &
$0.2020^{+0.00410}_{-0.00433}$ &
$0.1748^{+0.00850}_{-0.00781}$ &
$0.1800^{+0.00941}_{-0.00788}$ \\

$a/R_\star$ & Scaled semi-major axis &
$12.21^{+0.42}_{-0.40}$ &
$17.76^{+0.65}_{-0.61}$ &
$13.26^{+0.46}_{-0.45}$ &
$7.05^{+0.20}_{-0.19}$ \\

$i$ & Orbital inclination (deg) &
$88.41^{+0.76}_{-0.49}$ &
$88.89^{+0.45}_{-0.32}$ &
$86.43^{+0.29}_{-0.32}$ &
$83.93^{+0.40}_{-0.39}$ \\

$b$ & Impact parameter &
$0.34^{+0.11}_{-0.16}$ &
$0.34^{+0.10}_{-0.14}$ &
$0.75^{+0.05}_{-0.06}$ &
$0.74^{+0.03}_{-0.04}$ \\

$T_{14}$ & Transit duration (days) &
$0.08470^{+0.00297}_{-0.00326}$ &
$0.08612^{+0.00265}_{-0.00265}$ &
$0.07244^{+0.00434}_{-0.00378}$ &
$0.05246^{+0.00167}_{-0.00153}$ \\

\hline
\multicolumn{6}{c}{\textbf{Planetary Parameters}} \\
\hline
$M_p$ & Planet mass ($M_\oplus$) &
$159.04^{+30.96}_{-29.56}$ &
$224.95^{+29.58}_{-29.51}$ &
$194.78^{+20.95}_{-20.86}$ &
$668.74^{+41.56}_{-46.37}$ \\

$M_p$ & Planet mass ($M_J$) &
$0.50^{+0.09}_{-0.09}$ &
$0.71^{+0.09}_{-0.09}$ &
$0.61^{+0.07}_{-0.07}$ &
$2.10^{+0.13}_{-0.15}$ \\

$R_p$ & Planet radius ($R_\oplus$) &
$10.9^{+0.52}_{-0.52}$ &
$10.7^{+0.49}_{-0.47}$ &
$11.3^{+0.69}_{-0.65}$ &
$11.5^{+0.69}_{-0.62}$ \\

$R_p$ & Planet radius ($R_J$) &
$0.98^{+0.05}_{-0.05}$ &
$0.95^{+0.04}_{-0.04}$ &
$1.01^{+0.06}_{-0.06}$ &
$1.02^{+0.06}_{-0.06}$ \\

$\rho_p$ & Planet density (g\,cm$^{-3}$) &
$0.67^{+0.17}_{-0.14}$ &
$1.01^{+0.21}_{-0.18}$ &
$0.74^{+0.18}_{-0.14}$ &
$2.44^{+0.48}_{-0.42}$ \\

$a$ & Semi-major axis (AU) &
$0.03294^{+0.00037}_{-0.00038}$ &
$0.04007^{+0.00052}_{-0.00054}$ &
$0.03664^{+0.00041}_{-0.00041}$ &
$0.01913^{+0.00021}_{-0.00021}$ \\

$S$ & Instellation ($S_\oplus$) &
$49.3^{+8.63}_{-8.63}$ &
$17.7^{+3.38}_{-3.38}$ &
$51.8^{+8.79}_{-8.79}$ &
$156.7^{+26.9}_{-26.9}$ \\

$T_{\rm eq}$ & Equilibrium temperature (K) &
$738^{+32}_{-32}$ &
$571^{+27}_{-27}$ &
$747^{+32}_{-32}$ &
$986^{+42}_{-42}$ \\
\hline\hline
\end{tabular*}
\end{sidewaystable*}

\section{Discussion} \label{sec:discussion}

The four planets presented here span a mass range of $\sim0.5-2.1~M_J$ while orbiting a narrow range of early-M dwarf host stars, providing a useful snapshot of giant planet demographics in the low-mass stellar regime. Three of the systems--TOI-7189\,b, TOI-7265B\,b, and TOI-7393\,b--have bulk densities broadly comparable to that of Saturn (0.687~g~cm$^{-3}$), with $\rho_p = 0.67^{+0.17}_{-0.14}$, $1.01^{+0.21}_{-0.18}$, and $0.74^{+0.18}_{-0.14}$~g~cm$^{-3}$, respectively. Despite having very similar radii ($R_p \sim 0.95$--$1.01\,R_{\rm J}$), these planets span a mass range of $\sim0.50$--$0.71\,M_J$, suggesting varying heavy-element enrichment and internal structure among gas-rich planets forming around low-mass stars, even at similar present-day orbital separations and host-star properties. In contrast, TOI-7394B\,b is significantly more massive ($M_p = 2.10^{+0.13}_{-0.15}\,M_{\rm J}$) and dense ($\rho_p = 2.44^{+0.48}_{-0.42}\,\rm g\,cm^{-3}$), placing it in the super-Jupiter regime. Together, these systems populate the two mass regimes highlighted in previous GEMS results: TOI-7189\,b, TOI-7265B\,b, and TOI-7393\,b join the growing sample of Saturn-mass, Jupiter-sized gas giants \citep[e.g.,][]{fernandes2025, reji2025, sandoval2026}, while TOI-7394B\,b represents the higher-mass tail of a smaller number of super-Jupiters \citep{kanodia2025, kanodia2025fgk}.

\begin{figure*}
    \centering
    \includegraphics[width=2\columnwidth]{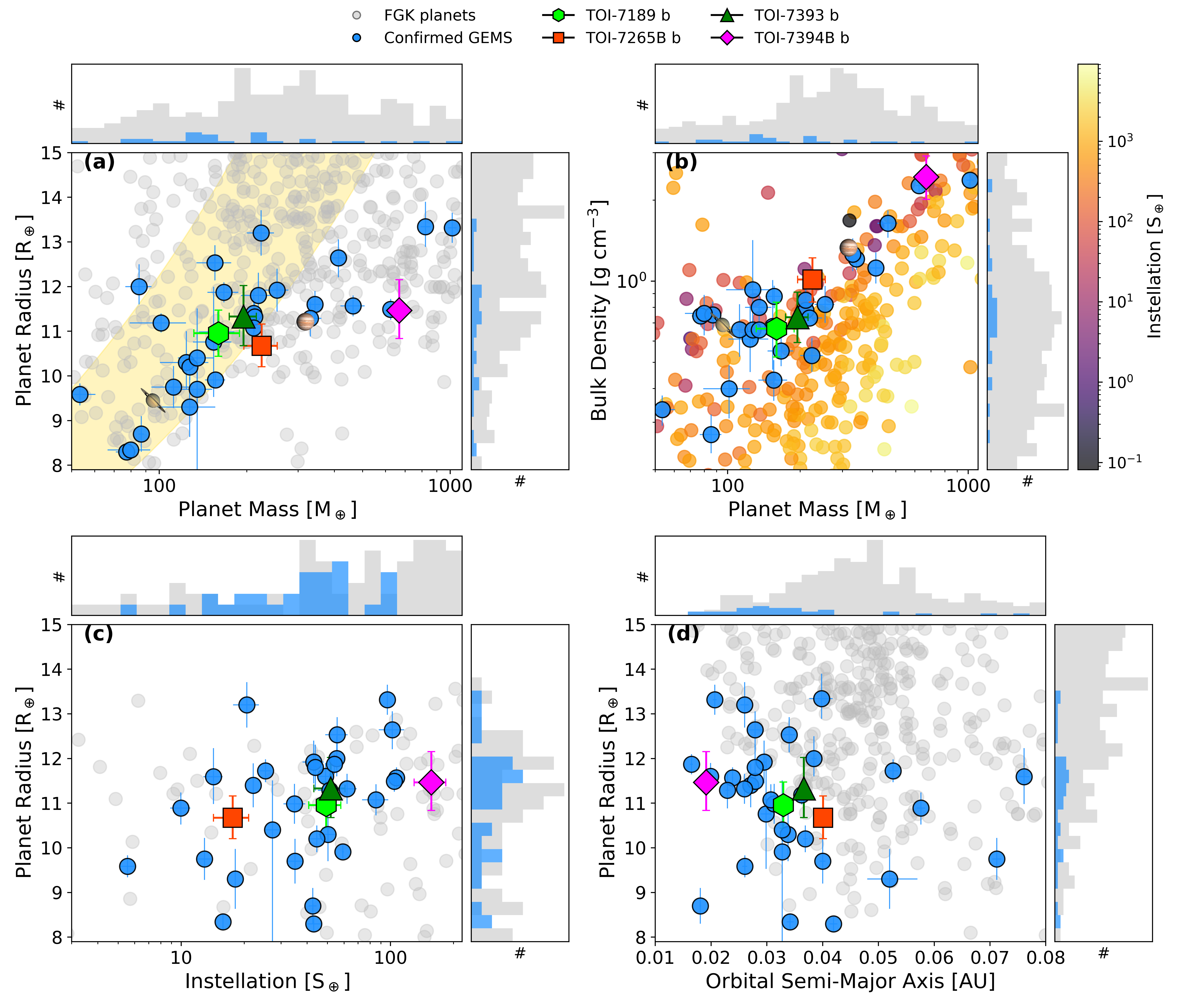}
\caption{
Multi-panel comparison of confirmed giant planets, highlighting TOI-7265B\,b, TOI-7189\,b, TOI-7393\,b, and TOI-7394B\,b. Blue points denote all confirmed GEMS planets with reported $1\sigma$ uncertainties from the NASA Exoplanet Archive, while light gray points show confirmed giant planets orbiting FGK stars for context. 
\textit{(a)} Planet mass versus radius, with the Saturn-density regime ($\rho = 0.3$--$0.9$~g~cm$^{-3}$) shaded in gold.
\textit{(b)} Planet mass versus bulk density.
\textit{(c)} Planetary instellation versus radius.
\textit{(d)} Orbital semi-major axis versus planet radius.
Marginal histograms along the top and right of each panel show the corresponding one-dimensional distributions of the FGK (gray) and GEMS (blue) samples after applying the same selection cuts as in the scatter plots. Inflated giant planets around FGK stars are predominantly found at high irradiation levels ($\gtrsim$160~$S_{\oplus}$), consistent with irradiation-driven radius inflation, whereas the GEMS planets occupy a lower-instellation regime. Three of the systems (TOI-7189\,b, TOI-7265B\,b, and TOI-7393\,b) lie within the locus of warm, Saturn-density planets commonly found around early-M dwarfs, while TOI-7394B\,b extends the GEMS population to higher masses and densities in the short-period super-Jupiter regime. Together, these systems illustrate the broad diversity of giant planets orbiting low-mass stars.
}
    \label{fig:mr}
\end{figure*}

In the mass--radius plane (Figure~\ref{fig:mr}), the three Saturn-mass planets occupy an intermediate region between Neptune-mass planets and the bulk of hot Jupiters orbiting FGK stars. For reference, we define the GEMS sample as confirmed transiting giant planets orbiting low-mass stars selected using the criteria $M_\star \leq 0.7M_\odot$, $T_{\rm eff} \leq 4000$K, orbital period $P \leq 10$days, and planetary radii $R_p \geq 8R_\oplus$, based on parameters compiled from the NASA Exoplanet Archive \citep{nea2025}. This region remains sparsely sampled for low-mass (late-K to mid-M dwarf) hosts, particularly at short orbital periods ($P < 5$~days) \citep{gan2023, bryant2023, glusman2025}. Their presence reinforces the emerging picture from the GEMS survey that warm, Saturn-density planets may represent a common outcome of giant planet formation around early-M dwarfs, while TOI-7394B\,b further demonstrates that substantially more massive gas giants can also form in this environment, albeit less frequently \citep{bryant2024, hotnisky2025}.

Despite their broadly similar sizes, the four planets exhibit a range of bulk densities. Among the three Saturn-mass planets, TOI-7265B\,b has a higher density, consistent with a greater degree of heavy-element enrichment, whereas TOI-7189\,b and TOI-7393\,b have lower densities indicative of more extended H/He-dominated envelopes, as expected from standard giant-planet interior models \citep{thorngren2016}. TOI-7394B\,b, while substantially more massive, exhibits the highest bulk density in the sample, consistent with the broader trend that more massive giant planets often contain larger total heavy-element masses and experience stronger interior compression at comparable radii \citep{miller2011, thorngren2016}.

At fixed radius, variations in bulk density primarily reflect differences in the total heavy-element content and envelope mass fraction, although the internal distribution and phase of these materials remain unconstrained \citep{fortney2007, thorngren2016}. The observed diversity among these systems therefore highlights the range of compositions accessible to giant planets forming around low-mass stars and underscores the importance of population-level studies for constraining their formation pathways.

\subsection{Formation and Evolution}

Population-level constraints on GEMS indicate that close-in giant planets around M dwarfs are intrinsically rare, with occurrence rates at the $\lesssim 0.1\%$ level \citep{gan2023, bryant2023, glusman2025}, roughly an order of magnitude lower than the $\sim1$--$2\%$ occurrence measured for FGK stars \citep{beleznay2022}. This scarcity is broadly consistent with theoretical expectations that lower-mass protoplanetary disks and longer core-growth timescales around M dwarfs inhibit the formation of sufficiently massive solid cores prior to gas-disk dispersal \citep{laughlin2004, kennedy2008, burn2021}. Observations of protoplanetary disks show that disk masses scale roughly linearly with stellar mass ($M_{\rm disk} \propto M_\star$), reducing the available solid reservoir for core growth around low-mass stars \citep{andrews2013, pascucci2016}.

Despite these challenges, the four systems presented here demonstrate that M-dwarf disks can produce a wide range of giant-planet outcomes. Three of the systems orbit metal-rich hosts, consistent with metallicity-enhanced core growth, while TOI-7393\,b suggests that runaway gas accretion may also occur in disks with lower inferred metallicity, although the stellar abundance measurements remain uncertain.

The diversity in planet masses motivates considering the Saturn-mass and super-Jovian regimes separately, as these outcomes likely probe different stages or efficiencies of runaway gas accretion \citep{kanodia2025}. While gravitational instability has been proposed as an alternative formation pathway for giant planets \citep{rafikov2005}, it is generally expected to operate at wide separations in massive disks and is therefore unlikely to be the dominant formation pathway for the close-in systems in this sample.

The large radii and relatively low bulk densities of the three Saturn-mass planets, compared to evolutionary models predicting radii for irradiated giant planets at similar equilibrium temperatures \citep[e.g.,][]{fortney2007, thorngren2016}, indicate that they possess substantial H/He envelopes. Because M dwarfs provide lower incident flux than FGK hosts at comparable orbital separations, irradiation-driven radius inflation is expected to be less efficient in these systems; their observed radii therefore likely reflect intrinsic structural properties rather than strong inflation mechanisms. Envelope masses of this magnitude are difficult to reproduce through purely hydrostatic envelope growth in classical core-accretion models, suggesting that these planets likely entered the runaway gas-accretion regime, although the precise onset of runaway remains uncertain and may occur at higher core masses depending on accretion history and opacity evolution \citep[e.g.,][]{helled2023}. 

However, their final masses indicate that runaway accretion did not proceed long enough to produce Jovian-mass planets. Around low-mass stars, several mechanisms may limit the efficiency or duration of runaway growth. Protoplanetary disks around M dwarfs are observed to be less massive on average, reducing the gas supply available during the runaway phase \citep{andrews2013, pascucci2016}, while lower viscous accretion rates may further throttle gas delivery onto forming planets \citep{bitsch2015, mordasini2012}. In addition, continued accretion of solids during envelope growth can delay cooling and contraction of the atmosphere, postponing runaway accretion until late stages of disk evolution \citep{piso2014, venturini2017}. If runaway begins only shortly before gas-disk dispersal---typically after $\sim$1--5 Myr \citep{alexander2014}---planets may accrete substantial envelopes yet fail to reach Jupiter mass before the gas reservoir is removed.

These processes likely act together to regulate the final masses of gas giants forming around low-mass stars, where reduced disk masses and shorter effective accretion windows may limit prolonged runaway growth \citep{laughlin2004}. Recent GEMS discoveries have demonstrated that gas giants with masses comparable to Saturn and Jupiter can form around M dwarfs spanning a wide range of stellar masses. Comparative studies further indicate that warm Jupiters around M dwarfs exhibit masses and bulk densities similar to those orbiting FGK stars, suggesting that broadly similar formation pathways may operate across stellar-mass regimes \citep{kanodia2025fgk, kanodia2025}.

In this context, the Saturn-mass planets in our sample are consistent with scenarios in which runaway gas accretion begins but is halted by disk evolution before the planet reaches Jupiter mass, though, as previously detailed, the formation pathways of Saturn-mass planets remain uncertain. The presence of the more massive TOI-7394B\,b in the same stellar-mass regime further suggests that, under favorable disk conditions, runaway accretion around M dwarfs can proceed efficiently enough to produce Jovian or super-Jovian planets. This result is broadly consistent with population-level analyses showing that while super-Jupiter mass planets do occur around M dwarfs, the most massive gas giants are comparatively rare around these hosts \citep{delamer2024, hotnisky2025}.

Recent theoretical work emphasizes that the onset and outcome of runaway gas accretion may be more complex than predicted by classical core-accretion models. Continued accretion of solids during envelope growth can delay cooling and contraction of the envelope, shifting the onset of runaway gas accretion to higher planetary masses and making the final outcome sensitive to disk properties and accretion history \citep{alibert2014, helled2023}. The diversity of planet masses observed here may therefore reflect variations in disk mass, metallicity, and gas-disk lifetime across otherwise similar low-mass stellar hosts. Observational studies of the growing GEMS sample also suggest that giant planets orbiting M dwarfs preferentially occur around metal-rich host stars, consistent with expectations from the core-accretion paradigm \citep{fischer2005}, and supported by trends emerging within the growing GEMS sample.

All four planets were most plausibly assembled beyond the snow line, where solid material is abundant, and subsequently migrated inward through disk-driven migration to their present short-period orbits \citep{baruteau2014, bitsch2019}, as in-situ formation of giant planets at these orbital separations is strongly disfavored by core-accretion models. This implies that the observed short-period giant planet population represents a migration-modified subset of the underlying giant-planet population formed at larger orbital separations. As a result, present-day orbital architectures are not direct tracers of the formation locations of these planets. Post-formation evolution may further shape their present-day densities. Because M dwarfs are less luminous than FGK stars, irradiation-driven radius inflation is expected to be less efficient in these systems, potentially allowing differences in heavy-element enrichment and core mass to play a larger role in determining the observed structural diversity \citep{fortney2007, thorngren2016}.

\subsection{Host-Star Metallicity in the GEMS Population}

\begin{figure*}
    \centering
    \includegraphics[width=\textwidth]{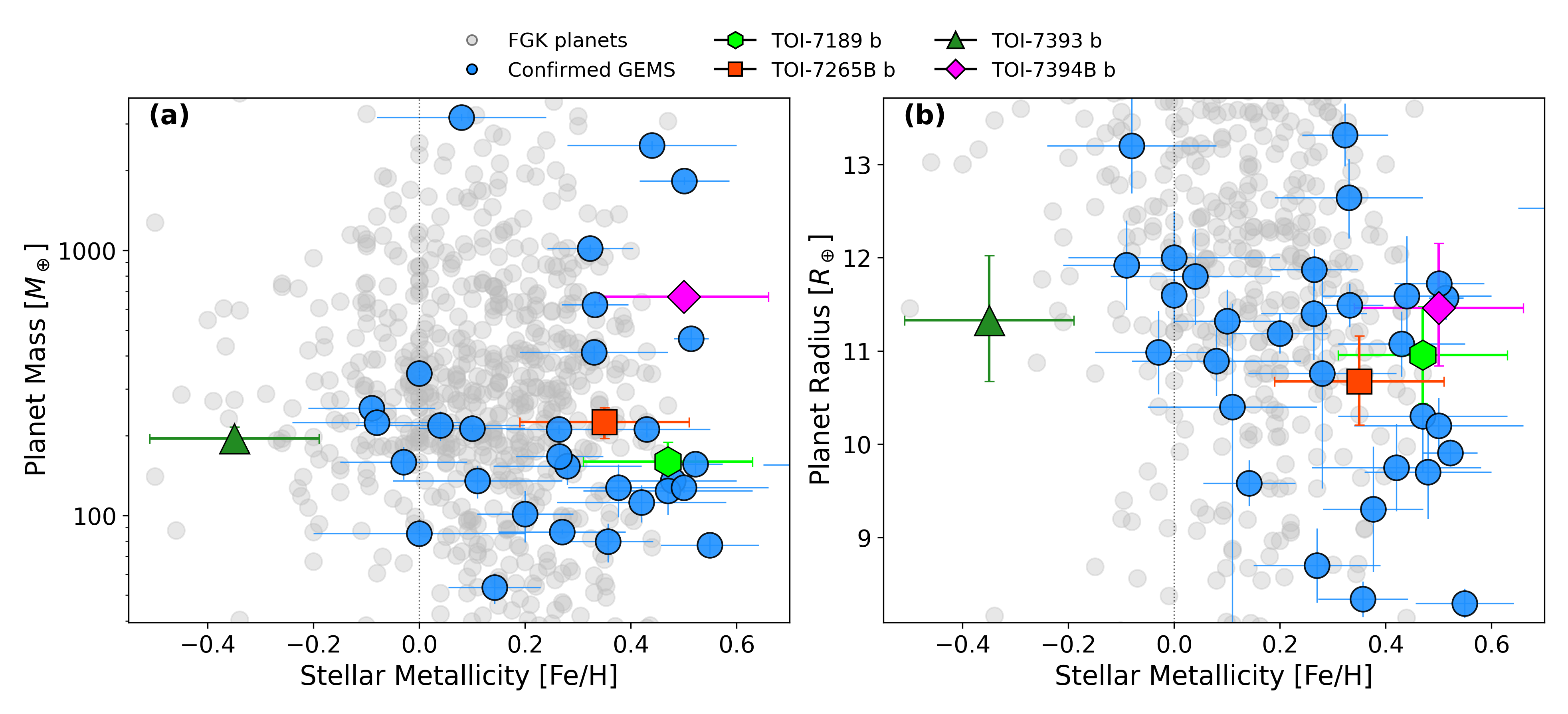}
\caption{
Planet mass (left) and planet radius (right) as a function of host-star metallicity for confirmed GEMS planets.
Blue points denote the confirmed GEMS population, while highlighted symbols mark TOI-7265B\,b, TOI-7189\,b, TOI-7393\,b, and TOI-7394B\,b. Most GEMS giant planets orbit metal-rich M dwarfs, consistent with the well-known metallicity dependence of giant-planet formation. Three of the systems presented here follow this trend, while TOI-7393\,b orbits a significantly metal-poor host, extending the GEMS population to lower metallicities and demonstrating that giant planets can occasionally form even in comparatively metal-poor environments.
}
    \label{fig:metallicity}
\end{figure*}

The four systems presented here span a broad range of host-star metallicities and planet masses while orbiting a relatively narrow range of early-M dwarfs, providing new insight into the role of stellar properties in shaping giant planet formation around low-mass stars. Three of the host stars (TOI-7265B, TOI-7189, and TOI-7394B) have stellar masses of $\sim0.5$--$0.6\,M_\odot$ and comparatively high inferred metallicities. While these stars lie near the upper end of the metallicity distribution within the confirmed GEMS sample (Figure~\ref{fig:metallicity}), we note that the apparent upper bound is partly imposed by the parameter coverage of the \texttt{SpecMatch} spectral library, which limits the range of recoverable metallicities and therefore should not be interpreted as a physical metallicity ceiling. Within these uncertainties, these hosts represent favorable conditions for giant planet formation around low-mass stars, consistent with the well-established correlation between stellar metallicity and giant-planet occurrence observed for FGK hosts \citep{gonzalez1997, santos2004, fischer2005}.

The super-solar metallicities of TOI-7189, TOI-7265B, and TOI-7394B are consistent with enhanced disk solid surface densities that promote rapid core growth and facilitate runaway gas accretion, producing gas-rich planets ranging from Saturn mass to super-Jovian masses. The presence of the massive TOI-7394B\,b further demonstrates that, under favorable conditions, giant planet formation around early-M dwarfs can proceed efficiently to produce Jovian and super-Jovian planets.

On the other hand, TOI-7393 orbits a substantially metal-poor host ($\mathrm{[Fe/H]} = -0.35 \pm 0.16$), placing it among the most metal-poor M-dwarf hosts known to harbor a gas giant, and among the more metal-poor giant-planet hosts across FGKM stars. Within the current GEMS sample, only four other systems have sub-solar metallicities, all of which lie in the relatively narrow range $-0.1 \lesssim \mathrm{[Fe/H]} < 0$. TOI-7393 therefore extends the metallicity range of known GEMS host stars toward lower values. Its $\sim0.6\,M_J$ planet demonstrates that giant planets can form around metal-poor M dwarfs, although such systems remain relatively uncommon in the present sample. This interpretation is further supported by its kinematics ($TD/D \sim 0.6$), which place it in the thin/thick-disk transition and are consistent with an older, more metal-poor stellar population \citep{haywood2008}.

Taken together, these four systems outline a formation sequence in which increasing disk solid content and overall formation efficiency promote progressively more massive outcomes. Saturn-mass planets appear to represent the typical endpoint of giant planet formation around early-M dwarfs, while the production of Jovian and super-Jovian planets likely requires the most metal-rich and massive disks. TOI-7394B\,b may represent the upper extreme of this pathway, whereas TOI-7393--consistent with an older, metal-poor population based on its kinematics--serves as a reminder that metallicity, while strongly predictive, does not uniquely determine giant-planet formation outcomes and that alternative pathways may operate in this sample.

\subsection{Prospects for Atmospheric Characterization}

To assess the suitability of these systems for atmospheric follow-up, we estimated the Transmission Spectroscopy Metric (TSM) following \citet{kempton2018}. We obtain TSM values of 71 for TOI-7189\,b, 49 for TOI-7265B\,b, 66 for TOI-7393\,b, and 24 for TOI-7394B\,b. The relatively large TSM values for TOI-7189\,b and TOI-7393\,b are driven by their low bulk densities and correspondingly extended atmospheres, making them favorable targets for transmission spectroscopy with facilities such as \textit{JWST} and the upcoming \textit{Ariel} mission \citep{tinetti2018, gardner2023}. In contrast, the substantially higher mass and surface gravity of TOI-7394B\,b likely suppress its atmospheric scale height despite its elevated equilibrium temperature. 

These systems are particularly valuable in the context of comparative atmospheric surveys. A primary goal of \textit{Ariel} is to characterize the atmospheres of $\sim$1000 exoplanets spanning a wide range of planetary and stellar properties in order to identify population-level trends linking atmospheric composition to planet formation and evolution \citep{tinetti2018}. GEMS remain significantly underrepresented within current atmospheric samples, especially in the warm Saturn- to Jupiter-mass regime explored here. Collectively, these four systems span a broad range of irradiation levels, surface gravities, and host-star metallicities, making them useful benchmarks for comparitive studies of atmospheric composition and giant-planet formation around low-mass stars. These systems also complement ongoing GEMS \textit{JWST} efforts targeting warm giant planets orbiting M-dwarfs, including the broader atmospheric survey and population-level analysis presented by \citet{canas2026}.

\section{Conclusions} \label{sec:conclusion}

We present the confirmation and characterization of four short-period transiting giant planets orbiting nearby low-mass stars: TOI-7189\,b, TOI-7265B\,b, TOI-7393\,b, and TOI-7394B\,b. Using a combination of \textit{TESS} photometry, ground-based transit observations, high-resolution spectroscopy, and precise radial velocity measurements, we confirm all four systems as gas-rich planets orbiting early- to mid-M dwarfs.

Our joint photometric and radial velocity modeling yields the following planetary parameters:
\begin{itemize}
    \item \textbf{TOI-7189\,b:} $R_p = 0.98^{+0.05}_{-0.05}\,R_{\rm J}$, $M_p = 0.50^{+0.09}_{-0.09}\,M_{\rm J}$, corresponding to a bulk density of $\rho_p = 0.67^{+0.17}_{-0.14}$~g\,cm$^{-3}$, similar to Saturn.

    \item \textbf{TOI-7265B\,b:} $R_p = 0.95^{+0.04}_{-0.04}\,R_{\rm J}$, $M_p = 0.71^{+0.09}_{-0.09}\,M_{\rm J}$, corresponding to a bulk density of $\rho_p = 1.01^{+0.21}_{-0.18}$~g\,cm$^{-3}$.

    \item \textbf{TOI-7393\,b:} $R_p = 1.01^{+0.06}_{-0.06}\,R_{\rm J}$, $M_p = 0.61^{+0.07}_{-0.07}\,M_{\rm J}$, implying a relatively low bulk density of $\rho_p = 0.74^{+0.18}_{-0.14}$~g\,cm$^{-3}$ and an extended H/He envelope.

    \item \textbf{TOI-7394B\,b:} $R_p = 1.02^{+0.06}_{-0.06}\,R_{\rm J}$, $M_p = 2.10^{+0.13}_{-0.15}\,M_{\rm J}$, yielding a bulk density of $\rho_p = 2.44^{+0.48}_{-0.42}$~g\,cm$^{-3}$ and placing it in a higher-gravity regime.
\end{itemize}
Together, these planets span a factor of $\sim4$ in mass and a wide range of bulk densities and surface gravities, indicating substantial diversity in envelope mass fractions and heavy-element enrichment among gas-rich planets orbiting low-mass stars.

All four planets orbit relatively massive ($M_* \sim 0.5$--$0.6\,M_\odot$) early-M dwarfs, but they span a broad range of host-star metallicities. Three systems (TOI-7189, TOI-7265B, and TOI-7394B) orbit super-solar metallicity stars, consistent with the well-established correlation between stellar metallicity and giant planet occurrence. In contrast, TOI-7393 orbits a substantially metal-poor host ($\mathrm{[Fe/H]} \approx -0.35$), making it the most metal-poor giant-planet host currently known within the GEMS sample. Its existence demonstrates that while metallicity strongly enhances the likelihood of giant planet formation, it is not a strict prerequisite. Disk mass, lifetime, or stochastic variations in accretion efficiency may occasionally compensate for reduced solid content, enabling gas-giant formation even in metal-poor environments.

Within the context of the GEMS survey and related occurrence-rate studies, short-period giant planets around M dwarfs are intrinsically rare, with occurrence rates $\lesssim 0.1\%$ \citep{kanodia2024a, glusman2025}. The predominantly Saturn-mass outcomes in this sample are consistent with formation in lower-mass protoplanetary disks, where limited gas reservoirs and shorter disk lifetimes likely truncate runaway growth before planets reach Jovian masses \citep{fischer2005, mordasini2012, helled2023}. The $2.10\,M_J$ planet TOI-7394B\,b represents the high-mass extreme of this distribution and demonstrates that, under especially favorable conditions, early-M dwarfs are capable of producing Jovian and even super-Jovian planets. Together, TOI-7189\,b, TOI-7265B\,b, TOI-7393\,b, and TOI-7394B\,b provide important benchmarks for testing theories of planet formation, migration, and atmospheric evolution in the low-mass stellar regime.

\begin{acknowledgments}
We thank the anonymous referee for a careful and constructive review that improved the clarity and quality of this manuscript.

This research has been partially supported by the NASA United States Contributions to the Ariel Preparatory Science Program (USCAPS) under grant number NASA-80NSSC25K0184.  The research has been done as part of the  NASA Ariel Science Center (NASC) managed by JPL for NASA and operated at Caltech/IPAC in support of NASA's participation in the ESA mission Ariel. 
This research has made use of the Exoplanet Follow-up Observation Program (ExoFOP; DOI: 10.26134/ExoFOP5) website, which is operated by the California Institute of Technology, under contract with the National Aeronautics and Space Administration under the Exoplanet Exploration Program. 

This work utilized (i) the NASA Exoplanet Archive, operated by Caltech under contract with NASA through the Exoplanet Exploration Program; (ii) the SIMBAD database, operated at CDS, Strasbourg, France; (iii) NASA’s Astrophysics Data System Bibliographic Services; and (iv) data from the Two Micron All Sky Survey (2MASS), a joint project of the University of Massachusetts and IPAC/Caltech, funded by NASA and the NSF.

This work has made use of the VALD database, operated at Uppsala University, the Institute of Astronomy RAS in Moscow, and the University of Vienna. The authors acknowledge the use of Open AI ChatGPT to help format tables. 

These results are based on observations obtained with the Habitable-zone Planet Finder Spectrograph on the HET. We acknowledge support from NSF grants AST-1006676, AST-1126413, AST-1310885, AST-1310875,  ATI-2009889, ATI-2009554, ATI-2009982, AST-2108512, AST-2108801 and the NASA Astrobiology Institute (NNA09DA76A) in the pursuit of precision radial velocities in the NIR. The HPF team also acknowledges support from the Heising-Simons Foundation via grant 2017-0494.

The Hobby-Eberly Telescope is a joint project of the University of Texas at Austin, the Pennsylvania State University, Ludwig-Maximilians-Universität München, and Georg-August Universität Gottingen. The HET is named in honor of its principal benefactors, William P. Hobby and Robert E. Eberly. The HET collaboration acknowledges the support and resources from the Texas Advanced Computing Center. We thank the Resident astronomers and Telescope Operators at the HET for the skillful execution of our observations with HPF. We would like to acknowledge that the HET is built on Indigenous land. Moreover, we would like to acknowledge and pay our respects to the Carrizo \& Comecrudo, Coahuiltecan, Caddo, Tonkawa, Comanche, Lipan Apache, Alabama-Coushatta, Kickapoo, Tigua Pueblo, and all the American Indian and Indigenous Peoples and communities who have been or have become a part of these lands and territories in Texas, on Turtle Island.

The Center for Exoplanets and Habitable Worlds is supported by Penn State and its Eberly College of Science.

WIYN is a joint facility of the University of Wisconsin--Madison, Indiana University, NSF’s NOIRLab, the Pennsylvania State University, Purdue University, the University of California--Irvine, and the University of Missouri. We thank the
NEID Queue Observers and WIYN Observing Associates for their skillful execution of our NEID observations.

The authors are honored to be permitted to conduct astronomical research on Iolkam Du’ag (Kitt Peak), a mountain with particular significance to the Tohono O’odham. Data presented herein were obtained at the WIYN Observatory from telescope time allocated to NN-EXPLORE through the scientific partnership of NASA, the NSF, and NOIRLab. Some of the observations in this paper made use of the NN-EXPLORE Exoplanet and Stellar Speckle Imager (NESSI). NESSI was funded by the NASA Exoplanet Exploration Program and the NASA Ames Research Center. NESSI was built at the Ames Research Center by Steve B. Howell, Nic Scott, Elliott P. Horch, and Emmett Quigley.

CIC acknowledges support by NASA Headquarters through an appointment to the NASA Postdoctoral Program at the Goddard Space Flight Center, administered by ORAU through a contract with NASA, and support from NASA under award number 80GSFC24M0006.

Resources supporting this work were provided by the NASA Scientific Computing project through the NASA Center for Climate Simulation (NCCS) at Goddard Space Flight Center. This content is solely the responsibility of the authors and does not necessarily represent the views of the NCCS.

\end{acknowledgments}

\facilities{TESS, RBO:0.6m, PC:1m, HET (HPF), WIYN (NEID, NESSI), Gaia}

\software{
\texttt{ArviZ} \citep{arviz2019},
\texttt{astropy} \citep{astropy2013},
\texttt{AstroImageJ} \citep{collins2016},
\texttt{celerite2} \citep{celerite2018},
\texttt{EXOFASTv2} \citep{eastman2019},
\texttt{exoplanet} \citep{foremanmackey2021},
\texttt{HPF-SpecMatch} \citep{stefansson2020},
\texttt{HPF-SERVAL} \citep{hpfserval},
\texttt{HxRGproc} \citep{ninan2018},
\texttt{iPython} \citep{ipython2007},
\texttt{lightkurve} \citep{lightkurve2018},
\texttt{matplotlib} \citep{matplotlib2007},
\texttt{metamorphosis}\citep{duque2022},
\texttt{pyastrotools} \citep{pyastro2023}, 
\texttt{pymc3} \citep{salvatier2016}, 
\texttt{scipy} \citep{scipy2020},
\texttt{TGLC} \citep{han2023},
\texttt{Theano} \citep{theano2016}
}

\bibliography{main}{}
\bibliographystyle{aasjournalv7}

\begin{table*}[ht!]
\centering
\renewcommand{\arraystretch}{0.9}
\caption{
Photometric noise, dilution, rotational variability, and radial velocity jitter parameters for each instrument, observing date, and \textit{TESS} sector. All \textit{TESS} photometric jitter values are reported in ppt.
}
\label{tab:appendix_jitter}
\begin{tabular*}{\textwidth}{@{\extracolsep{\fill}}llcccc}
\hline\hline
Parameter & Description & TOI-7189 & TOI-7265B & TOI-7393 & TOI-7394B \\
\hline
\multicolumn{6}{c}{\textbf{Radial Velocity Jitter and Offsets}} \\
\hline
$\sigma_{\rm HPF}$ & HPF RV jitter (m\,s$^{-1}$) &
$37.09^{+17.10}_{-15.51}$ &
$35.48^{+17.43}_{-14.67}$ &
$25.76^{+16.55}_{-14.78}$ &
$44.05^{+28.80}_{-22.80}$ \\

$\sigma_{\rm NEID_{pre}}$ & NEID Pre 2025 Dec 9 RV jitter (m\,s$^{-1}$) &
-- & -- & -- &
$54.41^{+31.31}_{-35.31}$ \\

$\sigma_{\rm NEID_{post}}$ & NEID Post 2025 Dec 9 RV jitter (m\,s$^{-1}$) &
-- & -- &
$19.81^{+26.57}_{-13.92}$ &
$60.59^{+28.36}_{-38.66}$ \\

$\gamma_{\rm HPF}$ & HPF RV offset (m\,s$^{-1}$) &
$-71.29^{+14.35}_{-14.75}$ &
$-13.01^{+13.29}_{-13.73}$ &
$-38.82^{+12.80}_{-12.63}$ &
$-42.36^{+27.29}_{-29.34}$ \\

$\gamma_{\rm NEID_{pre}}$ & NEID Pre 2025 Dec 9 RV offset (m\,s$^{-1}$) &
-- & -- & -- &
$-223.08^{+61.09}_{-58.67}$\\

$\gamma_{\rm NEID_{post}}$ & NEID Post 2025 Dec 9 RV offset (m\,s$^{-1}$) &
-- & -- &
$-28.68^{+16.24}_{-15.73}$ &
$-491.18^{+75.47}_{-74.97}$ \\

\hline
\multicolumn{6}{c}{\textbf{TESS Photometric Jitter (ppt)}} \\
\hline
$\sigma_{\rm TESS,S15}$ & Sector 15 &
-- &
$10.92^{+0.37}_{-0.35}$ &
-- &
$3.82^{+0.01}_{-0.00}$ \\

$\sigma_{\rm TESS,S16}$ & Sector 16 &
-- &
$12.00^{+0.56}_{-0.62}$ &
$2.64^{+0.10}_{-0.09}$ &
$4.60^{+0.13}_{-0.12}$ \\

$\sigma_{\rm TESS,S22}$ & Sector 22 &
-- & -- &
$2.05^{+0.07}_{-0.07}$ &
$2.74^{+0.14}_{-0.14}$ \\

$\sigma_{\rm TESS,S23}$ & Sector 23 &
-- & -- &
$1.91^{+0.09}_{-0.08}$ &
-- \\

$\sigma_{\rm TESS,S24}$ & Sector 24 &
-- & -- &
$2.86^{+0.12}_{-0.11}$ &
$4.28^{+0.12}_{-0.12}$ \\

$\sigma_{\rm TESS,S49}$ & Sector 49 &
-- & -- &
$4.37^{+0.06}_{-0.06}$ &
$5.54^{+0.20}_{-0.20}$ \\

$\sigma_{\rm TESS,S50}$ & Sector 50 &
-- & -- &
$3.53^{+0.05}_{-0.05}$ &
$7.83^{+0.11}_{-0.11}$ \\

$\sigma_{\rm TESS,S51}$ & Sector 51 &
-- & -- & -- &
$7.45^{+0.12}_{-0.12}$ \\

$\sigma_{\rm TESS,S55}$ & Sector 55 &
-- &
$18.00^{+0.23}_{-0.23}$ &
-- & -- \\

$\sigma_{\rm TESS,S56}$ & Sector 56 &
-- &
$33.75^{+0.23}_{-0.23}$ &
-- & -- \\

$\sigma_{\rm TESS,S75}$ & Sector 75 &
-- &
$29.63^{+0.20}_{-0.20}$ &
-- & -- \\

$\sigma_{\rm TESS,S76}$ & Sector 76 &
-- &
$31.72^{+0.22}_{-0.22}$ &
$4.25^{+0.03}_{-0.03}$ &
$9.56^{+0.15}_{-0.15}$ \\

$\sigma_{\rm TESS,S77}$ & Sector 77 &
-- & -- &
$7.83^{+0.08}_{-0.07}$ &
$11.98^{+0.12}_{-0.11}$ \\

$\sigma_{\rm TESS,S78}$ & Sector 78 &
-- & -- & -- &
$14.97^{+0.15}_{-0.15}$ \\

$\sigma_{\rm TESS,S80}$ & Sector 80 &
$0.03^{+0.26}_{-0.03}$ &
-- & -- & -- \\

$\sigma_{\rm TESS,S82}$ & Sector 82 &
-- &
$38.41^{+0.27}_{-0.27}$ &
-- &
$12.22^{+0.09}_{-0.08}$ \\

$\sigma_{\rm TESS,S83}$ & Sector 83 &
-- &
$34.58^{+0.25}_{-0.24}$ &
-- & -- \\

\hline
\multicolumn{6}{c}{\textbf{Ground-based Photometric Jitter (ppt)}} \\
\hline
$\sigma_{\rm Keeble,20250531}$ & Keeble (2025 May 31) &
$0.06^{+0.87}_{-0.06}$ &
-- & -- & -- \\

$\sigma_{\rm Pomona,20250722}$ & Pomona (2025 July 22) &
$9.25^{+0.24}_{-0.23}$ &
-- & -- & -- \\

$\sigma_{\rm RBO,20250727}$ & RBO (2025 July 27) &
-- &
$3.28^{+0.90}_{-0.94}$ &
-- & -- \\

$\sigma_{\rm RBO,20250818}$ & RBO (2025 August 18) &
-- & -- &
$0.036^{+0.42}_{-0.03}$ &
-- \\

$\sigma_{\rm Pomona,20251006}$ & Pomona (2025 October 6) &
-- &
$8.24^{+0.39}_{-0.37}$ &
-- & -- \\

$\sigma_{\rm RBO,20251011}$ & RBO (2025 October 11) &
$15.97^{+2.50}_{-2.08}$ &
-- & -- & -- \\

$\sigma_{\rm Keeble,20251014}$ & Keeble (2025 October 14) &
$4.28^{+0.84}_{-0.92}$ &
-- & -- & -- \\

$\sigma_{\rm RBO,20251121}$ & RBO (2025 November 21) &
-- &
$10.68^{+1.08}_{-0.92}$ &
-- & -- \\

$\sigma_{\rm Keeble,20260116}$ & Keeble (2026 January 16) &
-- & -- &
$3.00^{+0.55}_{-0.57}$ &
-- \\

$\sigma_{\rm Keeble,20260120}$ & Keeble (2026 January 20) &
-- & -- & -- &
$0.036^{+0.38}_{-0.03}$ \\

$\sigma_{\rm RBO,20260204}$ & RBO (2026 February 4) &
-- & -- &
$0.042^{+0.49}_{-0.04}$ &
-- \\

$\sigma_{\rm RBO,20260219}$ & RBO (2026 February 19) &
-- & -- & -- &
$0.57^{+0.80}_{-0.22}$ \\

\hline
\multicolumn{6}{c}{\textbf{TESS Dilution Factors}} \\
\hline
$D_{\rm TESS,S15}$ & Sector 15 &
-- &
$1.14^{+0.15}_{-0.15}$ &
-- &
$0.796^{+0.01}_{-0.01}$ \\

$D_{\rm TESS,S16}$ & Sector 16 &
-- &
$1.12^{+0.13}_{-0.12}$ &
$1.18^{+0.10}_{-0.09}$ &
$1.01^{+0.10}_{-0.10}$ \\

$D_{\rm TESS,S22}$ & Sector 22 &
-- & -- &
$1.24^{+0.10}_{-0.09}$ &
$1.02^{+0.09}_{-0.10}$ \\

$D_{\rm TESS,S23}$ & Sector 23 &
-- & -- &
$1.08^{+0.09}_{-0.08}$ &
-- \\

$D_{\rm TESS,S24}$ & Sector 24 &
-- & -- &
$1.13^{+0.10}_{-0.09}$ &
$0.84^{+0.09}_{-0.09}$ \\

$D_{\rm TESS,S49}$ & Sector 49 &
-- & -- &
$1.27^{+0.10}_{-0.10}$ &
$1.22^{+0.11}_{-0.12}$ \\

$D_{\rm TESS,S50}$ & Sector 50 &
-- & -- &
$1.00^{+0.09}_{-0.08}$ &
$1.05^{+0.10}_{-0.11}$ \\

$D_{\rm TESS,S51}$ & Sector 51 &
-- & -- & -- &
$1.06^{+0.10}_{-0.11}$ \\

$D_{\rm TESS,S55}$ & Sector 55 &
-- &
$1.09^{+0.10}_{-0.10}$ &
-- & -- \\

$D_{\rm TESS,S56}$ & Sector 56 &
-- &
$1.12^{+0.11}_{-0.10}$ &
-- & -- \\

$D_{\rm TESS,S75}$ & Sector 75 &
-- &
$1.27^{+0.10}_{-0.09}$ &
-- & -- \\

$D_{\rm TESS,S76}$ & Sector 76 &
-- &
$1.11^{+0.15}_{-0.14}$ &
$0.61^{+0.05}_{-0.05}$ &
$1.05^{+0.10}_{-0.10}$ \\

$D_{\rm TESS,S77}$ & Sector 77 &
-- & -- &
$1.15^{+0.10}_{-0.09}$ &
$0.91^{+0.09}_{-0.09}$ \\

$D_{\rm TESS,S78}$ & Sector 78 &
-- & -- & -- &
$1.30^{+0.12}_{-0.13}$ \\

$D_{\rm TESS,S80}$ & Sector 80 &
$1.11^{+0.05}_{-0.05}$ &
-- & -- & -- \\

$D_{\rm TESS,S82}$ & Sector 82 &
-- &
$1.27^{+0.10}_{-0.09}$ &
-- &
$1.12^{+0.10}_{-0.11}$ \\

$D_{\rm TESS,S83}$ & Sector 83 &
-- &
$0.98^{+0.10}_{-0.10}$ &
-- & -- \\

\hline\hline
\end{tabular*}
\end{table*}

\end{document}